\documentclass[a4paper,12pt]{article}
\pdfoutput=1
\usepackage{jheppub}

\title{\boldmath Unravelling AdS$_3$/CFT$_2$ near the boundary}

\abstract{We study correlation functions of spectrally-flowed vertex operators in bosonic string theory on $\text{AdS}_3\times X$ in the path integral formalism. By restricting the path integral to only include worldsheets which live near the asymptotic boundary, we compute correlation functions of spectrally-flowed vertex operators and find a precise agreement with the perturbative correlators in the recently-proposed dual CFT at all orders in conformal perturbation theory. We thus provide highly nontrivial evidence for the bulk/boundary duality.}

\author[a]{Bob Knighton,}
\author[b]{Vit Sriprachyakul}
\affiliation[a]{Department of Applied Mathematics \& Theoretical Physics, University of Cambridge,\\
Wilberforce Road, Cambridge CB3 0WA, United Kingdom}
\affiliation[b]{Institut f\"ur Theoretische Physik,
ETH Z\"urich,\\
Wolfgang-Pauli-Strasse 27,
8093 Z\"urich, Switzerland}
\emailAdd{rik23@cam.ac.uk}
\emailAdd{vsriprachyak@phys.ethz.ch}

\usepackage{bm} 
\usepackage[dvipsnames]{xcolor}
\definecolor{green_maf}{RGB}{28, 166, 46}
\definecolor{blue_mrg}{RGB}{12, 143, 145}
\definecolor{detail}{RGB}{110,110,110}
\usepackage{amsmath} 
\usepackage{nccmath} 
\usepackage{amssymb} 
\usepackage{amsthm} 
\usepackage{mathtools} 
\usepackage[utf8]{inputenc} 
\usepackage{braket}
\usepackage{enumerate}
\usepackage[many]{tcolorbox}
\usepackage[inline]{enumitem}
\usepackage{comment}
\usepackage{soul} 
\usepackage{tikz}
\usepackage{csquotes}

\newtcolorbox{empheqboxed}{colback=gray!30, 
 colframe=white,
 width=\textwidth,
 sharpish corners,
 top=-2mm, 
 bottom=0pt
}

\hypersetup{
		pdfencoding=unicode,
		colorlinks=true,
		urlcolor=Maroon,
		linkcolor=RoyalBlue,
		citecolor=Maroon,
		pdfstartview=FitH,
		linktocpage=true
}

\usetikzlibrary{calc,decorations.markings}
\usetikzlibrary{decorations.pathmorphing}
\usetikzlibrary{shapes,backgrounds}
\usetikzlibrary{fadings}
\usetikzlibrary{cd}
\usetikzlibrary{decorations.pathreplacing,calligraphy}
\usetikzlibrary{hobby}

\tikzset{
	partial ellipse/.style args={#1:#2:#3}{
		insert path={+ (#1:#3) arc (#1:#2:#3)}
	}
}

\tikzset{
  every overlay node/.style={
    draw=black,fill=white,rounded corners,anchor=north west,
  },
}

\tikzfading
[
  name=fade out,
  inner color=transparent!0,
  outer color=transparent!100
]

\newif\ifdetails
\detailstrue

\def\be{\begin{equation}}
\def\ee{\end{equation}}

\begin{document}

\maketitle

\section{Introduction}

The AdS/CFT correspondence relates string theories on Anti-de Sitter spacetimes to conformal field theories on their boundary. Due to the strong/weak nature of this duality, examples for which both sides are analytically solvable are invaluable for a deeper understanding of the holographic principle. A specifically nice set of such dualities are the $\text{AdS}_3/\text{CFT}_2$ dualities, for which both the bulk worldsheet theories and boundary CFTs are under much better control than in higher dimensions. The simplification of the worldsheet theory lies in the accidental isometry $\text{AdS}_3\cong\text{SL}(2,\mathbb{R})$, so that the worldsheet CFT is described by a Wess-Zumino-Witten (WZW) model on $\text{SL}(2,\mathbb{R})$, assuming that the background is supported purely by NS-NS flux \cite{Balog:1988jb,Petropoulos:1989fc,Hwang:1990aq,Henningson:1991jc,Gawedzki:1991yu,Bars:1995mf,Evans:1998qu,Giveon:1998ns,Kutasov:1999xu,Maldacena:2000hw,Maldacena:2000kv,Maldacena:2001km}.

\begin{figure}
\centering
\begin{tikzpicture}
\begin{scope}[xscale = 1.4]
\draw[thick, latex-latex] (-4.5,0) -- (4.5,0);
\draw[thick, -latex] (-3.8,0) -- (-3.8,5.3);
\node[above left] at (-3.8,5.3) {$V(\varphi)$};
\node[above right] at (4.5,0) {$\varphi$};
\draw[smooth, domain=-2.3:4.3, thick] plot (\x,{2^(-\x)});
\fill[gray, opacity = 0.2] (0,0) -- (-4.3,0) -- (-4.3,5) -- (0,5);
\fill[gray, opacity = 0.05] (0,0) -- (4.3,0) -- (4.3,5) -- (0,5);
\node[above] at (-2,-0.6) {\footnotesize Nonperturbative};
\node[below] at (-2,-0.4) {\footnotesize regime};
\node[above] at (2,-0.6) {\footnotesize Perturbative};
\node[below] at (2,-0.4) {\footnotesize regime};
\draw[thick, dashed] (0,0) -- (0,5);
\node[above] at (-2,5) {\footnotesize Deep interior};
\node[above] at (2,5) {\footnotesize Asymptotic boundary};
\end{scope}
\end{tikzpicture}
\caption{The linear dilaton $\varphi$ of the dual CFT is deformed by an exponential wall. In the regime $\varphi\to\infty$ (corresponding to the boundary of $\text{AdS}_3$) this wall can be regarded as a perturbation, while in the opposite regime $\varphi\to-\infty$ (corresponding to the deep interior of $\text{AdS}_3$) it must be treated non-perturbatively.}
\label{fig:strong-weak}
\end{figure}
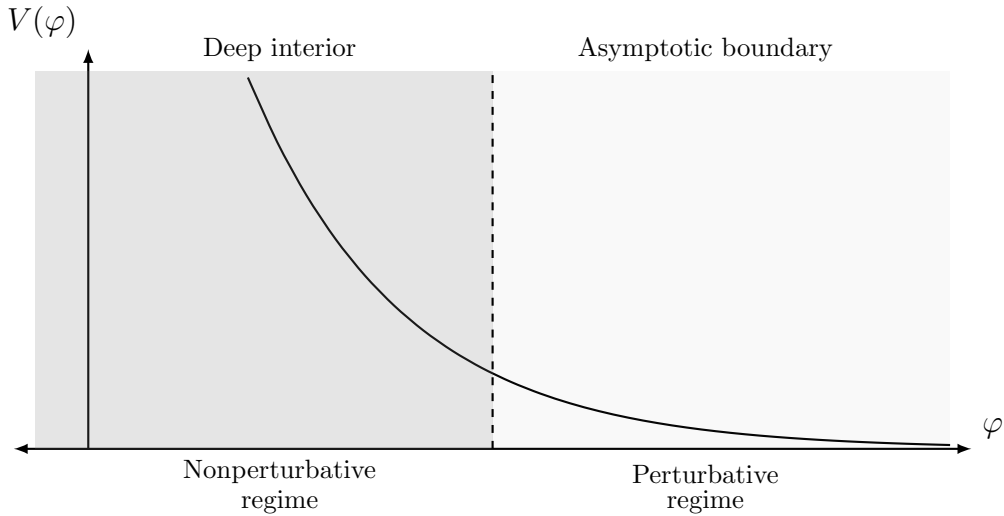

Recently, a family of holographic CFTs have been proposed to be dual to pure NS-NS $\text{AdS}_3\times X$ backgrounds \cite{Eberhardt:2019qcl,Balthazar:2021xeh,Martinec:2021vpk,Eberhardt:2021vsx}. The proposed field theories are modeled by a symmetric product orbifold of a non-compact CFT, deformed by a marginal twist-2 field. The duality we will be most interested in in this paper is that of \cite{Eberhardt:2021vsx}, which relates \textit{bosonic} string theory on $\text{AdS}_3\times X$ to the symmetric orbifold of a linear dilaton theory, perturbed by an exponential `wall'. Specifically, the proposal is\footnote{The dual CFT is strictly speaking only defined in the large $K$ limit, and so should be thought of as dual to \textit{perturbative} string theory.}
\begin{equation}\label{eq:string-duality}
\begin{gathered}
\text{Bosonic string theory on AdS}_3\times X\\
\Longleftrightarrow\\
\text{Sym}^K(\mathbb{R}_{\varphi}\times X)+\mu\int\sigma_{2,\alpha}\,,
\end{gathered}
\end{equation}
where $\mathbb{R}_{\varphi}$ is a linear dilaton of background charge $\mathcal{Q}=\sqrt{2}(3-k)/\sqrt{k-2}$ and $\sigma_{2,\alpha}$ is a marginal field in the twist-2 sector whose profile is shown in Figure \ref{fig:strong-weak}. In this way, the dual CFT is morally analogous to the definition of Liouville theory as a deformation of a linear dilaton, except that the exponential wall now lives in the twisted sector of an orbifold group.

In \cite{Eberhardt:2021vsx}, evidence for this duality was provided in the form of a matching of spectrally-flowed 2- and 3-point functions in $\text{AdS}_3$ with twisted-sector correlators in the dual CFT up to fourth order in $\mu$. Matching of the 4-point functions at zeroth order in $\mu$ was found in \cite{Dei:2022pkr}.\footnote{The $\text{AdS}_3$ correlation functions in question were computed in \cite{Dei:2021xgh,Dei:2021yom,Dei:2022pkr} by solving the $\mathfrak{sl}(2,\mathbb{R})_k$ Ward identities and the Knizhnik–Zamolodchikov equations explicitly for a large number of spectral flows $w_i$ and proposing a general formula for all $w_i$. No proof of the general formulae in \cite{Dei:2021xgh,Dei:2021yom,Dei:2022pkr} is known, except for in the case of 3-point functions \cite{Bufalini:2022toj}.} The matching of the AdS$_3$ and dual CFT correlation functions is highly nontrivial, and required the conjectural relations between various analytic data of certain branched holomorphic covering maps $\gamma:\text{S}^2\to\text{S}^2$. However, matching $n$-point functions at all orders in $\mu$ is still an open problem.

The primary bottleneck in matching the two sides of \eqref{eq:string-duality} is the technical difficulty of calculating correlation functions in $\text{AdS}_3$ string theory. While correlation functions of the $\text{SL}(2,\mathbb{R})$ WZW model have been discussed at great length in the literature,\footnote{See \cite{Teschner:1997ft,Teschner:1999ug,Giribet:1999ft,Giribet:2000fy,Maldacena:2001km,Giribet:2001ft,Ribault:2005ms,Giribet:2005ix,Hikida:2007tq,Iguri:2007af,Baron:2008qf,Iguri:2009cf,Cagnacci:2013ufa,Giribet:2015oiy,Hikida:2020kil,Eberhardt:2019ywk,Eberhardt:2020akk,Dei:2021xgh,Dei:2021yom,Dei:2022pkr,Bufalini:2022toj} for a non-exhaustive list of references.} a full solution is still an open problem. Particularly difficult is the calculation of correlators of spectrally-flowed vertex operators in the so-called $x$-basis, which are still not known in closed form. The complication is two-fold. From the operator formalism, since spectrally-flowed states in $\text{AdS}_3$ live in non-highest-weight representations of $\mathfrak{sl}(2,\mathbb{R})_k$, their correlation functions obey very complicated Ward identities, which have only been solved analytically in very special cases \cite{Eberhardt:2019ywk,Eberhardt:2020akk,Dei:2021xgh,Dei:2021yom,Dei:2022pkr,Bufalini:2022toj}. From the path integral, the difficulty is that spectrally-flowed vertex operators do not admit obvious representations in terms of the fundamental fields of the worldsheet CFT, and so it is difficult to construct an integrand in the Polyakov path integral.

In \cite{Knighton:2023mhq,Hikida:2023jyc}, progress was made toward matching correlation functions of the dual CFT from the worldsheet in full generality. In \cite{Knighton:2023mhq}, it was shown how the the structure of the correlators in the CFT at order $\mu^m$ in perturbation theory can be reproduced by the path integral of $\text{AdS}_3$ string theory. The key ingredient was an explicit form of spectrally-flowed vertex operators in the worldsheet CFT. The resulting path integral was shown to localize to an integral over exactly $m$ points on the boundary of $\text{AdS}_3$, reflecting the integrated deformation $\int\sigma_{2,\alpha}$ in the dual CFT. However, this computation was schematic, and the form of the integrand of the path integral was not determined. Independently, the authors of \cite{Hikida:2023jyc} were able to use CFT technology developed in the context of the $\mathbb{H}_3^+$/Liouville duality \cite{Hikida:2007tq,Hikida:2008pe,Hikida:2020kil,Hikida:2023jyc} to show how the analytic forms of the dual CFT correlators can be derived from the $\text{AdS}_3$ path integral, although did not provide an explanation for the localization of the worldsheet path integral.

\vspace{1cm}

The goal of this paper is to provide a first principles derivation of the perturbative CFT correlation functions from the $\text{AdS}_3$ path integral at tree level in the worldsheet genus expansion. This is achieved by considering an effective theory of $\text{AdS}_3$ string theory near the asymptotic boundary. In this limit, the worldsheet theory simplifies, and it is possible to compute correlation functions in a Coulomb-gas-like formalism \cite{Knighton:2023mhq}. By a careful analysis of the path integral measure, as well as the integral over moduli space $\mathcal{M}_{0,n}$ on the worldsheet, we precisely reproduce the symmetric orbifold answer perturbatively in $\mu$.

The fact that we can reproduce the full perturbative expansion in $\mu$ by only considering a near-boundary approximation in $\text{AdS}_3$ reflects the fact that the perturbative expansion in $\mu$ is not the full boundary CFT answer. Just like in Liouville theory, the fact that the marginal operator in the symmetric orbifold is not normalizable means that perturbation theory breaks down and there are non-perturbative effects that must be taken into account. These effects come from the region of finite $\varphi$ in Figure \ref{fig:strong-weak}, which holographically correspond to contributions from the deep interior of $\text{AdS}_3$. Thus, the matching of the perturbative CFT correlators from the near-boundary theory on $\text{AdS}_3$ should be thought of as a tree-level proof of the slightly weaker duality:
\begin{equation}\label{eq:string-duality-weak}
\begin{gathered}
\text{Bosonic string theory on AdS}_3\times X\text{ near the boundary}\\
\Longleftrightarrow\\
\text{Sym}^K(\mathbb{R}_{\varphi}\times X)+\mu\int\sigma_{2,\alpha}\text{ in the perturbative regime}\,.
\end{gathered}
\end{equation}
Although this subsector of the full $\text{AdS}_3/\text{CFT}_2$ duality fails to capture nonperturbative (in $\mu$) effects, we view the work in this paper as highly nontrivial evidence for the full duality \eqref{eq:string-duality}.

\vspace{1cm}

This paper is organized as follows. In Sections \ref{sec:ads3-strings} and \ref{sec:cft-dual}, we review bosonic string theory on $\text{AdS}_3$ and its proposed CFT dual, respectively. While these sections contain no novel material, we include them for the sake of completeness, and expert readers should feel free to skip over them. In Section \ref{sec:correlators}, we explicitly compute correlation functions of spectrally-flowed states on $\text{AdS}_3$ in the path integral formalism, and demonstrate that they exactly agree with the perturbative correlation functions of the proposed dual CFT at all orders in conformal perturbation theory. We conclude in Section \ref{sec:discussion} with a brief summary of the main text and a list of potential future directions. We also include a technical appendix which computes the Jacobian of a certain change of variables necessary for the evaluation of the $\text{AdS}_3$ path integral.

\section{\boldmath Strings on \texorpdfstring{AdS$_3$}{AdS3}}\label{sec:ads3-strings}

The main goal of this paper is to compute string correlation functions in $\text{AdS}_3$ and compare them to dual CFT correlators. On the worldsheet side, this amounts to computing correlators of the form
\begin{equation}
\sum_{g=0}^{\infty}g_s^{2g-2}\int_{\mathcal{M}_{g,n}}\Braket{V_{m_1,\bar{m}_1,j_1}^{w_1}(x_1,z_1)\cdots V_{m_n,\bar{m}_n,j_n}^{w_n}(x_n,z_n)}\,,
\end{equation}
where the vertex operators $V_{m,\bar{m},j}^w(x,z)$ are associated with spectrally-flowed ground states in the worldsheet theory. In this section, we will review the basic notions of $\text{AdS}_3$ string theory necessary to compute the above correlators, and return to the actual computation later in Section \ref{sec:correlators}.

\subsection{The sigma model}

Euclidean $\text{AdS}_3$ in Poincar\'e coordinates has the metric tensor
\begin{equation}
\mathrm{d}s^2=\frac{k}{r^2}(\mathrm{d}r^2+\mathrm{d}\gamma\,\mathrm{d}\bar{\gamma})\,,
\end{equation}
where $r\in\mathbb{R}_+$ is a radial coordinate such that $r\to 0$ is the conformal boundary of $\text{AdS}_3$ and the complex coordinates $(\gamma,\bar{\gamma})$ parametrize the boundary, whose induced metric is flat everywhere except at the point $\gamma=\infty$. Here, $k$ is a length scale related to the $\text{AdS}$ radius $L$ by\footnote{Note that we set $\alpha'=1$. If we were to reinstate units, we would write $k=L^2/\alpha'$.}
\begin{equation}
k=L^2\,.
\end{equation}
In order to describe a consistent string background, we must also introduce a $B$-field
\begin{equation}
B=-\frac{k}{r^2}\mathrm{d}\gamma\wedge\mathrm{d}\bar{\gamma}\,.
\end{equation}
The inclusion of the $B$-field (as well as an internal CFT $X$ with central charge $c(X)=26-3k/(k-2)$) is required for the vanishing beta function of the worldsheet.\footnote{Alternatively, it can be seen as the Wess-Zumino term of the $\text{SL}(2,\mathbb{R})$ WZW model, which is guaranteed to be conformally invariant at the quantum level.}

To study string theory on this background, we find it convenient to introduce a scalar field $\Phi$ such that
\begin{equation}
r=r_0e^{-Q\Phi/2}\,,
\end{equation}
where $r_0$ is an arbitrary radius that sets a length scale. The constant $Q$ is related to the level $k$ by
\begin{equation}
Q=\sqrt{\frac{2}{k-2}}\,,
\end{equation}
and is chosen for later convenience. In these coordinates, it is possible to immediately write down the Polyakov action for the worldsheet theory, which will be a second-order action in the variables $(\Phi,\gamma,\bar{\gamma})$. However, for the purposes of this paper, we will be interested in the first-order formalism of $\text{AdS}_3$ strings, where one introduces Lagrange multipliers $\beta,\bar{\beta}$ of weight $(1,0)$ and $(0,1)$ respectively. The resulting action is \cite{deBoer:1998gyt,Giveon:1998ns,Kutasov:1999xu}
\begin{equation}\label{eq:wakimoto-action}
S=\frac{1}{2\pi}\int_{\Sigma}\left(\frac{1}{2}\partial\Phi\,\overline{\partial}\Phi+\beta\overline{\partial}\gamma+\bar{\beta}\partial\bar{\gamma}+\nu^{-1}\beta\bar{\beta}e^{-Q\Phi}-\frac{Q}{4}R\Phi\right)\,.
\end{equation}
The linear dilaton term $-QR\Phi/4$ is present due to a redefinition of the path integral measure that has been made upon introducing the Lagrange multipliers $\beta,\bar{\beta}$. The constant $\nu$ is a dimensionful parameter proportional to $r_0^{-2}$ (known as the `cosmological constant' in the literature), and can be set to any value by performing a shift in $\Phi$. While the shift $\Phi\to\Phi+\phi_0$ is not a symmetry of the above action, the combined shift
\begin{equation}\label{eq:shift-symmetry}
\Phi\to\Phi+\phi_0\,,\quad\nu\to e^{-Q\phi_0}\nu
\end{equation}
changes the action by a constant
\begin{equation}
S\to S-Q\phi_0(1-g)\,,
\end{equation}
as can be seen by the Gauss-Bonnet theorem. This implies that correlators of fields $\mathcal{O}_i$ carrying charge $q_i$ with respect to this symmetry satisfy
\begin{equation}
e^{\phi_0(\sum_{i}q_i+Q(1-g))}\Braket{\mathcal{O}_1\cdots\mathcal{O}_n}_{e^{Q\phi_0}\nu}=\Braket{\mathcal{O}_1\cdots\mathcal{O}_n}_{\nu}\,.
\end{equation}
Thus, the overall shift symmetry implies that correlators scale with $\nu$ like a power law:
\begin{equation}
\Braket{\mathcal{O}_1\cdots\mathcal{O}_n}_{\nu}\propto\nu^{-\frac{1}{Q}\sum_{i}q_i-(1-g)}\,.
\end{equation}

\subsubsection*{Quantization}

To quantize the $\text{AdS}_3$ sigma model, we can treat the action as a perturbation of a free field theory. In canonical quantization, the worldsheet fields can be treated to zeroth order as free fields satisfying the OPEs
\begin{equation}
\begin{gathered}
\Phi(z,\bar{z})\Phi(w,\bar{w})\sim-\log|z-w|^2\,,\\
\beta(z)\gamma(w)\sim-\frac{1}{z-w}\,,\quad\bar{\beta}(\bar{z})\bar{\gamma}(\bar{w})\sim-\frac{1}{\bar{z}-\bar{w}}\,.
\end{gathered}
\end{equation}
Here, we have explicitly included the (anti-)holomorphic behavior of the various fields. Later, we will be less careful and simply denote points on the worldsheet by a single coordinate $z$, and only re-introduce $\bar{z}$-dependence when there is ambiguity.

The fields above can be used to construct a left-moving current algebra generated by the currents
\begin{equation}
J^+=\beta\,,\quad J^3=-\frac{1}{Q}\partial\Phi+(\beta\gamma)\,,\quad J^-=-\frac{2}{Q}(\partial\Phi\,\gamma)+(\beta(\gamma\gamma))-k\partial\gamma\,,
\end{equation}
where $()$ denotes normal ordering. These currents satisfy the $\mathfrak{sl}(2,\mathbb{R})_k$ algebra\footnote{In Euclidean signature, the correct current algebra is technically $\mathfrak{sl}(2,\mathbb{C})_k$. However, it is common in the literature to be agnostic about the choice of real form and simply refer to the algebra as $\mathfrak{sl}(2,\mathbb{R})_k$ in both Euclidean and Lorenzian signature.}
\begin{equation}
\begin{gathered}
J^3(z)J^{\pm}(w)\sim\pm\frac{J^{\pm}(w)}{z-w}\,,\quad J^+(z)J^-(w)\sim\frac{k}{(z-w)^2}-\frac{2J^3(w)}{z-w}\,,\\
J^3(z)J^3(w)\sim-\frac{k/2}{(z-w)^2}\,.
\end{gathered}
\end{equation}
The zero modes of these currents generate the $\text{SL}(2,\mathbb{R})_L$ isometry of $\text{AdS}_3$ and their antiholomorphic counterparts generate the $\text{SL}(2,\mathbb{R})_R$ isometry.

\subsection{States and vertex operators}

Vertex operators in the $\text{AdS}_3$ sigma model fall into representations of the $\mathfrak{sl}(2,\mathbb{R})_k\times\overline{\mathfrak{sl}(2,\mathbb{R})}_k$ current algebra generated by the left- and right-moving currents $J^a,\bar{J}^a$. Focusing only on the left-movers, highest-weight representations of $\mathfrak{sl}(2,\mathbb{R})$ are spanned by states $\ket{m,j}$ with
\begin{equation}
J^3_0\ket{m,j}=m\ket{m,j}\,,\quad J^{\pm}_{0}\ket{m,j}=(m\pm j)\ket{m\pm 1,j}\,,
\end{equation}
and which are annihilated by the positive modes $J^a_n$, $n>0$. Such representations are labeled by the $\mathfrak{sl}(2,\mathbb{R})$ spin $j$ and the fractional part $\alpha$ of $m$. There are two cases of interest:
\begin{itemize}

    \item \textit{Discrete representations:} These are representations $\mathcal{D}^{+}_j$ and $\mathcal{D}^-_j$ which occur when there exists a highest/lowest weight state with respect to the raising, lowering operators $J^{\pm}_0$. Specifically,
    \begin{equation}
    \begin{split}
    \mathcal{D}^+_j&=\{\ket{m,j}\,,\,\,m=j,j+1,j+2,\ldots\}\,,\\
    \mathcal{D}^-_j&=\{\ket{m,j}\,,\,\,m=-j,-j-1,-j-2,\ldots\}\,,
    \end{split}
    \end{equation}
    along with all of their $J^a_{-n}$ descendants. As we will explain below, it is sufficient to only consider $\mathcal{D}^+_j$.

    \item \textit{Continuous representations:} These are representations $\mathcal{C}_{\alpha}^j$ spanned by states of the form
    \begin{equation}
    \mathcal{C}_j^{\alpha}=\{\ket{m,j}\,,\,\,m\in\mathbb{Z}+\alpha\}\,,
    \end{equation}
    along with all of their $J^a_{-n}$ descendants. These representations enjoy a reflection symmetry, and the representations $\mathcal{C}^{\alpha}_{j}$ and $\mathcal{C}^{\alpha}_{1-j}$ are equivalent.
    
\end{itemize}
The discrete representations $\mathcal{D}^{+}_j$ are unitary for $\frac{1}{2}<j<\frac{k-1}{2}$ and the continuous representations $\mathcal{C}_{\alpha}^j$ are unitary for $j\in\frac{1}{2}+i\mathbb{R}$. Including the right-moving representations as well, the spectrum of highest-weight representations takes the form\footnote{By the reflection symmetry $j\to 1-j$, we can restrict $j\in\frac{1}{2}+i\mathbb{R}_+$ for the continuous representations.}
\begin{equation}
\left(\int_{0}^{1}\mathrm{d}\alpha\int_{\frac{1}{2}+i\mathbb{R}_+}\mathrm{d}j\,\mathcal{C}^j_{\alpha}\otimes\overline{\mathcal{C}}^j_{\alpha}\right)\oplus\left(\int_{\frac{1}{2}}^{\frac{k-1}{2}}\mathrm{d}j\,\mathcal{D}^+_j\otimes\overline{\mathcal{D}}^+_j\right)\,.
\end{equation}
By the state-operator correspondence, we can introduce vertex operators $V_{m,\bar{m},j}$ for the state $\ket{m,j}_L\otimes\ket{\bar{m},j}_R$. The representation theory described above can be repackaged into the OPEs
\begin{equation}\label{eq:unflowed-rep-opes}
\begin{split}
J^3(z)V_{m,\bar{m},j}(y)&\sim\frac{mV_{m,\bar{m},j}(y)}{z-y}\,,\\
J^{\pm}(z)V_{m,\bar{m},j}(y)&\sim\frac{(m\pm j)V_{m\pm 1,\bar{m},j}(y)}{z-y}\,.
\end{split}
\end{equation}
The action of the right-moving currents is completely analogous.

In this paper, we will primarily be interested in the continuous representations $\mathcal{C}^{j}_{\alpha}$, as they play a more concrete role in the dual CFT.\footnote{As explained in \cite{Eberhardt:2021vsx}, discrete representations of $\mathfrak{sl}(2,\mathbb{R})_k$ appear as bound states or `LSZ poles' in the dual CFT, see \cite{Aharony:2004xn}.} Because of the equivalence between the representations with spins $j$ and $1-j$, the vertex operators should obey a reflection relation
\begin{equation}
V_{m,\bar{m},1-j}=R(m,\bar{m},1-j)V_{m,\bar{m},j}\,,
\end{equation}
where $R(m,\bar{m},j)$ is a reflection coefficient satisfying
\begin{equation}
R(m,\bar{m},1-j)R(m,\bar{m},j)=1\,.
\end{equation}
While we will mostly keep $R$ undetermined, it can be determined by requiring compatibility with the OPEs \eqref{eq:unflowed-rep-opes}. This is sufficient to determine $R$ up to a function $R_j$ depending purely on the spins as
\begin{equation}
R(m,\bar{m},j)=R_j\nu^{2j-1}(2j-1)\frac{\Gamma(m+j)\Gamma(-\bar{m}+j)\Gamma(1-2j)}{\Gamma(m+1-j)\Gamma(-\bar{m}+1-j)\Gamma(2j)}\,,
\end{equation}
where $R_j$ is constrained to satisfy the constraint $R_jR_{1-j}=1$. The power of $\nu$ is conventional, but is chosen so that the full vertex operator has definite transformation properties under the shift symmetry \eqref{eq:shift-symmetry}.

A convenient realization of the vertex operators $V_{m,\bar{m},j}$ in terms of the worldsheet free fields takes the form
\begin{equation}\label{eq:unflowed-operators}
\begin{split}
V_{m,\bar{m},j}=\,&e^{-Qj\Phi}\gamma^{-m-j}\bar{\gamma}^{-\bar{m}-j}\\
&\hspace{0.5cm}+R(m,\bar{m},j)\,e^{-Q(1-j)\Phi}\gamma^{-m-(1-j)}\bar{\gamma}^{-\bar{m}-(1-j)}\,.
\end{split}
\end{equation}
This expression is strictly speaking only valid in the limit of large $\Phi$ and the `true' vertex operators have a series of subleading corrections, as explained in \cite{Giveon:1998ns,Kutasov:1999xu,Aharony:2004xn,Iguri:2007af,Hikida:2000ry}. These operators satisfy the reflection property under $j\to 1-j$ and have worldsheet conformal weight
\begin{equation}
\Delta=\overline{\Delta}=\frac{j(1-j)}{k-2}=\frac{\frac{1}{4}+s^2}{k-2}\,,
\end{equation}
with $j=\frac{1}{2}+is$.

\subsubsection*{Spectrally-flowed states}

In addition to highest-weight representations, the algebra $\mathfrak{sl}(2,\mathbb{R})_k$ admits an infinite class of non-highest-weight representations. These are obtained by composing the highest-weight representations with the automorphism
\begin{equation}
\sigma^w(J^{\pm}_m)=J^{\pm}_{m\mp w}\,,\quad\sigma^w(J^3_m)=J^3_{m}+\frac{kw}{2}\delta_{m,0}\,,
\end{equation}
defined for any $w\in\mathbb{Z}$. The full string spectrum is generated by the spectrally-flowed images $\sigma^w(\mathcal{C}^j_{\alpha})$ and $\sigma^w(\mathcal{D}^{\pm}_j)$ of the highest-weight representations discussed above. Coincidentally, there is an equivalence
\begin{equation}\label{eq:discrete-equivalence}
\sigma^w(\mathcal{D}_j^-)=\sigma^{w-1}(\mathcal{D}^+_{\frac{k}{2}-j})\,,
\end{equation}
and so we can safely ignore the discrete representations $\sigma^w(\mathcal{D}^-_j)$ in the spectrum. That is, the full string spectrum takes the form
\begin{equation}
\bigoplus_{w\in\mathbb{Z}}\left[\left(\int_{0}^{1}\mathrm{d}\alpha\int_{\frac{1}{2}+i\mathbb{R}_+}\mathrm{d}j\,\sigma^w(\mathcal{C}^j_{\alpha})\otimes\sigma^w(\overline{\mathcal{C}}^j_{\alpha})\right)\oplus\left(\int_{\frac{1}{2}}^{\frac{k-1}{2}}\mathrm{d}j\,\sigma^w(\mathcal{D}^+_j)\otimes\sigma^w(\overline{\mathcal{D}}^+_j)\right)\right]\,.
\end{equation}
Focusing on $w\geq 0$, vertex operators $V_{m,\bar{m},j}^w$ satisfy the OPEs
\begin{equation}
\begin{split}
J^3(z)V_{m,\bar{m},j}^{w}(y)&\sim\frac{(m+kw/2)V_{m,\bar{m},j}(y)}{z-y}\,,\\
J^{\pm}(z)V_{m,\bar{m},j}^{w}(y)&\sim\frac{(m\pm j)V_{m\pm 1,\bar{m},j}(y)}{(z-y)^{1\pm w}}\,.
\end{split}
\end{equation}
Since $J^3_0$ ($\bar{J}^3_0$) reads off the conformal weight $h$ ($\bar{h}$) of the dual CFT operator, we have
\begin{equation}
h=m+\frac{kw}{2}\,,\quad\bar{h}=\bar{m}+\frac{kw}{2}\,.
\label{hhbar}
\end{equation}
Furthermore, the vertex operators $V_{m,\bar{m},j}^{w}$ have worldsheet conformal weight
\begin{equation}\label{eq:spectral-flow-conformal-weight-worldsheet}
\begin{split}
\Delta&=\frac{j(1-j)}{k-2}-hw+\frac{kw^2}{4}\,,\\
\overline{\Delta}&=\frac{j(1-j)}{k-2}-\bar{h}w+\frac{kw^2}{4}\,.
\end{split}
\end{equation}

The spectrally-flowed vertex operators $V_{m,\bar{m},j}^{w}$ do not admit a clean expression as an analytic function of the free fields. However, as was pointed out in \cite{Knighton:2023mhq}, one can formally express spectrally-flowed vertex operators as formal distributions in the fundamental fields and their derivatives, namely
\begin{equation}\label{eq:flowed-operators}
\begin{split}
V_{m,\bar{m},j}^{w}=\,&e^{(w/Q-Qj)\Phi}\bigg(\frac{\partial^w\gamma}{w!}\bigg)^{-m-j}\bigg(\frac{\overline{\partial}^w\bar{\gamma}}{w!}\bigg)^{-\bar{m}-j}\delta_w^{(2)}(\gamma)\\
&\hspace{1cm}+R(m,\bar{m},j)\,e^{(w/Q-Q(1-j))\Phi}\bigg(\frac{\partial^w\gamma}{w!}\bigg)^{-m-(1-j)}\bigg(\frac{\overline{\partial}^w\bar{\gamma}}{w!}\bigg)^{-\bar{m}-(1-j)}\delta_w^{(2)}(\gamma)\,.
\end{split}
\end{equation}
The novelty of this expression lies in the delta function operators
\begin{equation}
\delta_w^{(2)}(\gamma):=\delta(\gamma)\delta(\partial\gamma)\cdots\delta\left(\frac{\partial^{w-1}\gamma}{(w-1)!}\right)\delta(\bar{\gamma})\delta(\overline{\partial}\bar{\gamma})\cdots\delta\left(\frac{\overline{\partial}^{w-1}\bar{\gamma}}{(w-1)!}\right)\,.
\end{equation}
Intuitively, these operators constrain the path integral to fields $\gamma$ which have a zero of order $w$ at the insertion point of $V_{m,\bar{m},j}^w$.\footnote{See \cite{Frenkel:2006fy,Frenkel:2008vz} for a similar construction in the context of topological sigma models.} Just as in the case of the unflowed operators \eqref{eq:unflowed-operators}, one should think of equation \eqref{eq:flowed-operators} as defining nontrivial terms in an asymptotic expansion valid at large $\Phi$.

For ease of notation, we will define the combination
\begin{equation}
\Phi_{m,\bar{m},j}^w:=e^{(w/Q-Qj)\Phi}\bigg(\frac{\partial^w\gamma}{w!}\bigg)^{-m-j}\bigg(\frac{\overline{\partial}^w\bar{\gamma}}{w!}\bigg)^{-\bar{m}-j}\delta_w^{(2)}(\gamma)\,,
\end{equation}
so that the full vertex operators take the form
\begin{equation}
V_{m,\bar{m},j}^{w}=\Phi_{m,\bar{m},j}^w+R(m,\bar{m},j)\Phi_{m,\bar{m},1-j}^{w}\,.
\end{equation}

\subsubsection*{\boldmath The $x$ basis}

The vertex operators described above correspond to strings emitted from the asymptotic past of $\text{AdS}_3$. In Euclidean language, this means that the strings are emitted from the point $\gamma=\bar{\gamma}=0$ on the conformal boundary. In order to describe strings emitted from an arbitrary point $x$ on the boundary, we can use the translation operator $L^{\rm CFT}_{-1}=J^+_0=\beta_0$. That is, we define
\begin{equation}
V_{m,\bar{m},j}^{w}(x,z):=e^{xJ^+_0}V_{m,\bar{m},j}^{w}(z)e^{-xJ^+_0}\,.
\end{equation}
Using the identities
\begin{equation}
e^{x\beta_0}\Phi e^{-x\beta_0}=\Phi\,,\quad e^{x\beta_0}\gamma e^{-x\beta_0}=\gamma-x\,,
\end{equation}
we can readily write down the $x$ basis vertex operators as
\begin{equation}
V_{m,\bar{m},j}^{w}(x,z)=\Phi_{m,\bar{m},j}^{w}(x,z)+R(m,\bar{m},j)\Phi_{m,\bar{m},1-j}^{w}(x,z)\,,
\label{fullverads3}
\end{equation}
with
\begin{equation}
\Phi_{m,\bar{m},j}^{w}(x,z)=e^{(w/Q-Qj)\Phi}\bigg(\frac{\partial^w\gamma}{w!}\bigg)^{-m-j}\bigg(\frac{\overline{\partial}^w\bar{\gamma}}{w!}\bigg)^{-\bar{m}-j}\delta_w^{(2)}(\gamma-x)\,.
\label{leadingverads3}
\end{equation}

\subsection{Screening operators}

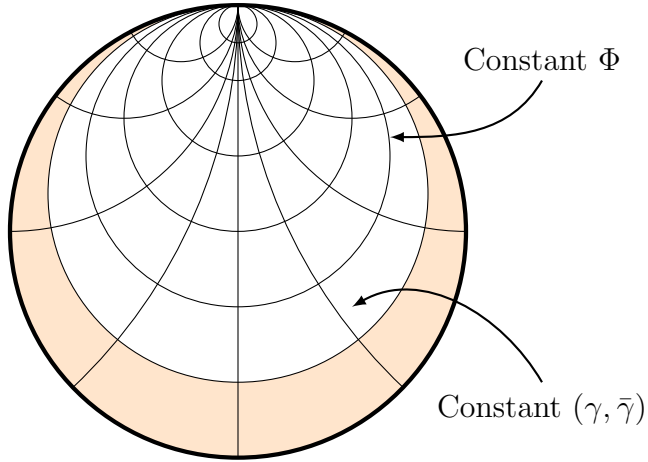
\begin{figure}
\centering
\begin{tikzpicture}
\fill[orange, opacity = 0.2] (0,0) circle (3);
\fill[white] (0,0.5) circle (2.5);
\draw (0,0.5) circle (2.5);
\draw (0,1) circle (2);
\draw (0,1.5) circle (1.5);
\draw (0,2) circle (1);
\draw (0,2.5) circle (0.5);
\draw (0,2.75) circle (0.25);
\draw (0,3) -- (0,-3);
\draw (0.75,3) [partial ellipse = 180:334:0.75 and 0.75];
\draw (-0.75,3) [partial ellipse = 0:-154:0.75 and 0.75];
\draw (1.5,3) [partial ellipse = 180:307:1.5 and 1.5];
\draw (-1.5,3) [partial ellipse = 0:-127:1.5 and 1.5];
\draw (3,3) [partial ellipse = 180:270:3 and 3];
\draw (-3,3) [partial ellipse = 0:-90:3 and 3];
\draw (7,3) [partial ellipse = 180:226.5:7 and 7];
\draw (-7,3) [partial ellipse = 0:-46.5:7 and 7];
\draw[ultra thick] (0,0) circle (3);
\draw[thick, -latex] (4,2) to[out = -120, in = 0] (2,1.25);
\node[above] at (4,2) {Constant $\Phi$};
\draw[thick, -latex] (4,-2) to[out = 120, in = 30] (1.5,-1);
\node[below] at (4,-2) {Constant $(\gamma,\bar{\gamma})$};
\end{tikzpicture}
\caption{Euclidean $\text{AdS}_3$ in Poincar\'e coordinates. The shaded orange region denotes the set $\Phi>\Phi_0$ for some large $\Phi_0$. Near the north pole $\gamma=\infty$, the coordinates are not defined.}
\label{fig:coordinate-patch}
\end{figure}

We are primarily interested in computing correlators in the asymptotic region of $\text{AdS}_3$ near the boundary, which is specified by $\Phi>\Phi_0$ for some large value of $\Phi_0$. However, as shown in Figure \ref{fig:coordinate-patch}, this subset does not include the entire conformal boundary, but rather misses the point at infinity. This is fundamentally due to working in Poincar\'e coordinates, which do not fully cover the conformal boundary. This problem must be remidied if we want to describe a CFT which lives on the full sphere $\text{S}^2$ instead of the complex plane $\mathbb{C}$.

One way around this would be to cover $\text{AdS}_3$ in two coordinate patches which together cover the full conformal boundary. For example, these coordinate patches could cover the sets $|\gamma|\leq 1/\varepsilon$ and $|\gamma|>1/\varepsilon$ for some $\varepsilon$. Taking $\varepsilon\to 0$ results in two patches: one which covers all finite values of $\gamma$ and one which covers $\gamma=\infty$. From the point of view of the worldsheet sigma model, including the second patch can be achieved by `allowing' $\gamma$ to diverge at certain points on the worldsheet. As was argued in \cite{Dei:2023ivl,Knighton:2023mhq,Hikida:2023jyc}, these divergences can be put in by hand by inserting vertex operators which transform trivially under all worldsheet symmetry currents $J^a$ but which have a simple pole with $\gamma$. 

Such a vertex operator can indeed be constructed and takes the form \cite{Giveon:2019gfk,Halder:2022ykw,Hikida:2023jyc,Knighton:2023mhq}\footnote{This screening operator is related to the so-called `secret' representations of \cite{Eberhardt:2019ywk}, as explained in \cite{Knighton:2023mhq}. It also shares many features with the so-caled $W$ field of the minimal tension worldsheet theory on $\text{AdS}_3\times\text{S}^3\times\mathbb{T}^4$ \cite{Dei:2020zui}.}
\begin{equation}
D=V_{m=\frac{k}{2},\bar{m}=\frac{k}{2},j=\frac{k-2}{2}}^{w=-1}\,.
\end{equation}
This operator has spacetime conformal weight $h=\bar{h}=0$ and worldsheet conformal weight $\Delta=\overline{\Delta}=1$. However, it does not exactly live in the trivial representation of $\mathfrak{sl}(2,\mathbb{R})_k$, but its integral does. Indeed, $D$ has trivial OPEs with $J^3$ and $J^+$, while its OPE with $J^-$ is a total derivative
\begin{equation}
J^-(z)D(w,\bar{w})\sim\partial_w(\cdots)\,,\quad \bar{J}^-(\bar{z})D(w,\bar{w})\sim\overline{\partial}_w(\cdots)\,.
\end{equation}
Thus, the nonlocal operator
\begin{equation}
\mathcal{O}=\int_{\Sigma}D
\end{equation}
is completely invisible to the $\mathfrak{sl}(2,\mathbb{R})_k$ current algebra. Furthermore, $D$ satisfies
\begin{equation}
\gamma(z)D(w,\bar{w})\sim\mathcal{O}\left(\frac{1}{z-w}\right)\,,\quad\bar{\gamma}(\bar{z})D(w,\bar{w})\sim\left(\frac{1}{\bar{z}-\bar{w}}\right)\,.
\end{equation}
For use in the path integral, it is helpful to write $D$ out in an explicit field-dependent form:
\begin{equation}
D=e^{-2\Phi/Q}\left|\oint\frac{\mathrm{d}z}{2\pi i}\,\gamma\right|^{-2(k-1)}\delta^{(2)}(\beta)\,.
\end{equation}
The contour integral should be taken around the insertion point of $D$, and can equivalently be thought of as reading off the residue of $\gamma$ at the pole enforced by the insertion of $D$. Again, we have defined the delta function operator
\begin{equation}
\delta^{(2)}(\beta):=\delta(\beta)\delta(\bar{\beta})\,.
\end{equation}
For the rest of the paper, we will refer to $\mathcal{O}$ as a \textit{screening operator}, keeping in tradition with the $\text{AdS}_3$ string theory literature \cite{Giribet:2001ft,Iguri:2007af}. We should also note that the operator $\mathcal{O}$ shares similarities with the interaction operator $\nu^{-1}\beta\bar{\beta}e^{-Q\Phi}$ in that they both have conformal weights $h=\bar{h}=0$, $\Delta=\overline{\Delta}=1$, have (almost) trivial OPEs with the currents $J^a$, and have a simple pole in their OPEs with $\gamma$ and $\bar{\gamma}$. However, they are crucially not equivalent operators (for one, they have different charges under $\partial\Phi$).

With the above definitions in mind, we propose that the correct prescription for computing correlators in the near-boundary region of Euclidean $\text{AdS}_3$ is to compute correlators in the free field theory with an arbitrary number of copies of $\mathcal{O}$ inserted. That is, we want to compute
\begin{equation}\label{eq:worldsheet-correlators-n-expansion}
\begin{split}
\sum_{N=0}^{\infty}\frac{p^N}{N!}&\Braket{\mathcal{O}^N \prod_{i=1}^{n}V_{m_i,\bar{m}_i,j_i}^{w_i}(x_i,z_i)}\\
&=\sum_{N=0}^{\infty}\frac{p^N}{N!}\int\mathrm{d}^2\lambda_1\cdots\mathrm{d}^2\lambda_N\Braket{D(\lambda_1,\bar{\lambda_1})\cdots D(\lambda_N,\bar{\lambda}_N)\prod_{i=1}^{n}V_{m_i,\bar{m}_i,j_i}^{w_i}(x_i,z_i)}\,.
\end{split}
\end{equation}
The factor of $1/N!$ is included since the positions of the poles of $\gamma$ are indistinguishable. We have also included a fugacity $p$ which counts the number of poles in $\gamma$. On dimensional grounds we have $p\propto\nu^{2-k}$, but we will keep it generic for now. Performing the sum over $N$, we see that including the screening operator $\mathcal{O}$ amounts to changing the action to
\begin{equation}
S=\frac{1}{2\pi}\int\left(\frac{1}{2}\partial\Phi\,\overline{\partial}\Phi+\beta\overline{\partial}\gamma+\bar{\beta}\partial\bar{\gamma}+\nu\beta\bar{\beta}e^{-Q\Phi}-\frac{Q}{4}R\Phi-pD\right)\,,
\label{fullaction}
\end{equation}
so that \eqref{eq:worldsheet-correlators-n-expansion} is reproduced by conformal perturbation theory in $p$. In Section \ref{sec:correlators} we will return to the problem of calculating the correlation functions in equation \eqref{eq:worldsheet-correlators-n-expansion} directly in the path integral formalism.

\section{The CFT dual}\label{sec:cft-dual}

Let us now turn our attention to the other side of the duality before actually computing worldsheet correlation functions in Section \ref{sec:correlators}. The CFT is based on a non-normalizable deformation of a symmetric orbifold \cite{Eberhardt:2021vsx,Dei:2022pkr}:
\begin{equation}
\text{Sym}(\mathbb{R}_{\mathcal{Q}}\times X)+\mu\int\sigma_{2,\alpha}\,,
\end{equation}
where $\mathbb{R}_{\mathcal{Q}}$ is a linear dilaton with background charge $\mathcal{Q}$ (which we specify below), and $\sigma_{2,\alpha}$ is a marginal operator in the twist-2 sector of the symmetric orbifold which carries momentum $\alpha$ with respect to the linear dilaton. We will review the details of this CFT as well as the computation of residues of poles in its correlators in the next few pages.

\subsection{The symmetric orbifold}

The dual CFT is based on a (non-normalizable) deformation of the symmetric product orbifold of $\mathbb{R}_{\mathcal{Q}}\times X$. The linear dilaton $\mathbb{R}_{\mathcal{Q}}$ is described by a scalar field $\varphi$ with action
\begin{equation}
S[\varphi]=\frac{1}{2\pi}\int_{B}\left(\frac{1}{2}\partial\varphi\,\overline{\partial}\varphi-\frac{\mathcal{Q}}{4}R\varphi\right)\,.
\end{equation}
Here, we have denoted by $B$ the conformal boundary $B\cong\partial\text{AdS}_3$. For the purpose of this paper, we are interested in the case $B\cong\text{S}^2$. The background charge $\mathcal{Q}$ is related to the level $k$ of the $\text{AdS}_3$ string theory by
\begin{equation}
\mathcal{Q}=Q-\frac{2}{Q}=-\frac{\sqrt{2}(k-3)}{\sqrt{k-2}}\,.
\end{equation}
The central charge of the theory is
\begin{equation}
c(\mathbb{R}_{\mathcal{Q}})=1+3\mathcal{Q}^2\,.
\end{equation}
The central charge of the internal CFT $X$ is determined by the consistency of the string background $\text{AdS}_3\times X$ to be
\begin{equation}
c(X)=26-\frac{3k}{k-2}\,,
\end{equation}
such that the total central charge is
\begin{equation}
c(\mathbb{R}_{\mathcal{Q}}\times X)=6k\,,
\end{equation}
as expected from the asymptotic symmetry algebra \cite{Giveon:1998ns}.
Primary states in the $\mathbb{R}_{\mathcal{Q}}\times X$ theory are given by tensor products of exponentials $e^{-q\varphi}$ with a state $\mathcal{O}_X$ of the internal CFT. We call $q$ the `momentum' of the linear dilaton $\varphi$, and the minus sign is chosen so that
\begin{equation}
\partial\varphi(z)\,e^{-q\varphi}(w)\sim\frac{q}{z-w}e^{-q\varphi}(w)\,.    
\end{equation}
The state $e^{-q\varphi}\otimes\mathcal{O}_X$ has conformal weight
\begin{equation}
h=\frac{q(\mathcal{Q}-q)}{2}+h_X\,,\quad\bar{h}=\frac{q(\mathcal{Q}-q)}{2}+\bar{h}_X\,.
\end{equation}

The symmetric product orbifold $\text{Sym}^K(\mathbb{R}_{\mathcal{Q}}\times X)$ is obtained by orbifolding $K$ copies of $\mathbb{R}_{\mathcal{Q}}\times X$ by the $S_K$ permutation symmetry interchanging them. As in any orbifold, the spectrum is organized into twisted and untwisted sectors labeled by conjugacy classes of the orbifold group. The single-cycle sectors are generated by permutations of the form $[(1\ldots w)]$ for $w\geq 1$. Given a vertex operator in $\mathbb{R}_{\mathcal{Q}}\times X$ obeying $h-\bar{h}\equiv 0\,\,(\text{mod }w)$, there exists an operator in the $w$-twisted sector with conformal weight
\begin{equation}
h_w=\frac{h}{w}+\frac{k(w^2-1)}{4w}\,,\quad\bar{h}_w=\frac{\bar{h}}{w}+\frac{k(w^2-1)}{4w}\,.
\end{equation}
The restriction on the allowed values of $(h,\bar{h})$ is such that $h_w-\bar{h}_w$ is integral, a necessity for locality of the theory. For the case of primary fields in $\mathbb{R}_{\mathcal{Q}}\times X$, we have
\begin{equation}\label{eq:symmetric-orbifold-conformal-weights}
\begin{split}
h_w&=\frac{q(\mathcal{Q}-q)}{2w}+\frac{h_X}{w}+\frac{k(w^2-1)}{4w}\,,\\
\bar{h}_w&=\frac{q(\mathcal{Q}-q)}{2w}+\frac{\bar{h}_X}{w}+\frac{k(w^2-1)}{4w}\,.
\end{split}
\end{equation}

The spectrum of single-cycle states can be recovered from the worldsheet theory on $\text{AdS}_3\times X$ by considering a state of the form $V_{m,\bar{m},j}^{w}\otimes\mathcal{O}_X$ and imposing the mass-shell condition $\Delta=\overline{\Delta}=1$. Using equation \eqref{eq:spectral-flow-conformal-weight-worldsheet}, the mass-shell equation is
\begin{equation}
\begin{split}
\frac{j(1-j)}{k-2}-h_ww+\frac{kw^2}{4}+h_X&=1\,,\\
\frac{j(1-j)}{k-2}-\bar{h}_ww+\frac{kw^2}{4}+\bar{h}_X&=1\,.
\end{split}
\end{equation}
Solving for $(h_w,\bar{h}_w)$ gives equation \eqref{eq:symmetric-orbifold-conformal-weights} if one makes the identification
\begin{equation}
q=Qj-\frac{1}{Q}\,.
\end{equation}
For strings living in the continuous representation $\sigma^w(\mathcal{C}^{j}_{\alpha})\otimes\sigma^w(\overline{\mathcal{C}}^{j}_{\alpha})$, we see that the worldsheet theory \textit{nearly} recovers the single-cycle spectrum of the symmetric product orbifold. However, there are two subtleties:
\begin{itemize}

	\item The worldsheet spin $j$ enjoys the reflection symmetry $j\to 1-j$. In terms of the dual CFT, this would naively imply the reflection symmetry
	\begin{equation}
	q\to \mathcal{Q}-q
	\end{equation}
	on the momenta of the linear dilaton $\varphi$.

	\item The spins in the continuous representations are taken to live on the strip $j\in\frac{1}{2}+i\mathbb{R}$. In terms of the momenta $q$, this condition would translate to
	\begin{equation}
	q\in\frac{\mathcal{Q}}{2}+i\mathbb{R}\,.
	\end{equation}

\end{itemize}
As was pointed out by the authors of \cite{Eberhardt:2019qcl}, both of these conditions can be satisfied if instead of a linear dilaton $\mathbb{R}_{\mathcal{Q}}$ one takes the symmetric orbifold of Liouville theory with background charge $\mathcal{Q}$. However, as was pointed out in \cite{Eberhardt:2021vsx}, the proposal of \cite{Eberhardt:2019qcl} does not produce the correct worldsheet $n$-point functions, and thus cannot be the correct description of the dual CFT.

\subsection{Perturbing the symmetric orbifold}

The proposal of \cite{Eberhardt:2021vsx} was to instead consider the symmetric orbifold $\text{Sym}(\mathbb{R}_{\mathcal{Q}}\times X)$ deformed by an operator of the form
\begin{equation}
S\to S+\mu\int\sigma_{2,\alpha}\,.
\end{equation}
Here, $\sigma_{2,\alpha}$ is a state in the $w=2$ twisted sector with linear dilaton momentum $\alpha$. This operator is classically marginal if $h(\sigma_{2,\alpha})=1$, which is solved by
\begin{equation}
\alpha=\frac{1}{Q}=\sqrt{\frac{k-2}{2}}\,.
\end{equation}
Just as in the case of Liouville theory, the operator $\sigma_{2,\alpha}$ creates an exponential potential in $\varphi$ of which states can scatter off. The result is that states with momenta $q$ and $\mathcal{Q}-q$ are no longer linearly independent. This leads to the identification of vertex operators with momentum $q$ and $\mathcal{Q}-q$, as predicted from the string theory. Focusing purely on the linear dilaton factor, states $\sigma_{w,q}$ in the single-cycle sector of the deformed theory are labeled by their twist $w$ and momentum $q$, and satisfy the reflection property
\begin{equation}
\sigma_{w,\mathcal{Q}-q}=R(h,\mathcal{Q}-q)\sigma_{w,\mathcal{Q}}\,,
\end{equation}
where $R(h,q)$ is a reflection coefficient which depends on the conformal weight $h$ of the state and satisfies the reflection property
\begin{equation}
R(h,q)R(h,\mathcal{Q}-q)=1\,.
\end{equation}

\subsection{Correlators in the dual CFT}\label{subsec:cft-correlators}

We consider correlation functions of the form
\begin{equation}\label{eq:goal-cft-correlator}
\Braket{\prod_{i=1}^{n}\sigma_{w_i,q_i}(x_i)}_{\mu}\,,
\end{equation}
where the subscript reminds us that we are computing these correlation functions in the deformed theory. Naively, this would be computed in conformal perturbation theory as
\begin{equation}\label{eq:naive-perturbative-answer}
\sum_{m=0}^{\infty}\frac{(-\mu)^{m}}{m!}\int\mathrm{d}^2\xi_1\cdots\mathrm{d}^2\xi_m\Braket{\prod_{i=1}^{n}\sigma_{w_i,q_i}(x_i)\prod_{\ell=1}^{m}\sigma_{2,\alpha}(\xi_\ell)}_{\mu=0}\,.
\end{equation}
However, like in Liouville theory, the non-normalizability of the operator $\sigma_{2,\alpha}$ means that naive perturbation theory is not valid.

The correct statement is as follows: the correlator \eqref{eq:goal-cft-correlator} scales with a power law in $\mu$, namely \cite{Eberhardt:2021vsx}
\begin{equation}
\Braket{\prod_{i=1}^{n}\sigma_{w_i,q_i}(x_i)}_{\mu,g}\propto\mu^{-(\sum_{i=1}^{n}q_i-\mathcal{Q}(1-g))/\alpha}\,,
\end{equation}
where $g$ is the genus of the contribution, determined by the order in the $1/K$ expansion. This correlator has poles when the exponent of $\mu$ is a non-negative integer, i.e. when
\begin{equation}
\sum_{i=1}^{n}q_i-\mathcal{Q}(1-g)=-\alpha m\,,\quad m\in\mathbb{Z}_{\geq 0}\,.
\end{equation}
Viewing \eqref{eq:goal-cft-correlator} as an analytic function in the sum $\sum_{i=1}^{n}q_i$ of the momenta, the residues of these poles will be given by \cite{Eberhardt:2021vsx}
\begin{equation}
\begin{split}
&\mathop{\text{Res}}_{\sum_{i}q_i-\mathcal{Q}(1-g)=-\alpha m}\Braket{\prod_{i=1}^{n}\sigma_{w_i,q_i}(x_i)}_{\mu}\\
&\hspace{2cm}=\frac{(-\mu)^{m}}{m!}\int\mathrm{d}^2\xi_1\cdots\mathrm{d}^2\xi_m\Braket{\prod_{i=1}^{n}\sigma_{w_i,q_i}(x_i)\prod_{\ell=1}^{m}\sigma_{2,\alpha}(\xi_\ell)}_{\mu=0}\,.
\end{split}
\end{equation}
Thus, while the form of the full interacting correlator \eqref{eq:goal-cft-correlator} is not known, by naive perturbation theory we can extract the residues of the poles in momentum space. Currently, this is all of the analytic data that can be explicitly calculated in the CFT proposal of \cite{Eberhardt:2021vsx}, and it is precisely the data we will recover below from the near-boundary limit of AdS$_3$ string theory.

\subsubsection*{Symmetric orbifold correlators}

Let us work out the naive perturbative answer \eqref{eq:naive-perturbative-answer} explicitly. For this, we need to know how to compute correlators of the form\footnote{In general, we could also consider nontrivial states in the seed CFT $X$. While we focus exclusively on states which are in the vacuum of $X$ here, the following discussion, as well as the calculations in Section \ref{sec:correlators}, generalize in a straightforward manner.}
\begin{equation}\label{eq:naive-single-m}
\Braket{\prod_{i=1}^{n}\sigma_{w_i,q_i}(x_i)\prod_{\ell=1}^{m}\sigma_{2,\alpha}(\xi_\ell)}_{\mu=0}\,.
\end{equation}
Such correlation functions can be computed using the covering space method \cite{Hamidi:1986vh,Lunin:2001ne}. This method begins by noting that the path integral computing the above correlator is an integral over fields which are not single-valued. In order to compute such a correlator, it is simpler to move to a covering space on which the fundamental fields lift to single-valued fields. Each covering space is realized by a surface $\Sigma$ along with a holomorphic map $\gamma:\Sigma\to\text{S}^2$ satisfying
\begin{equation}
\begin{split}
\gamma(z)\sim x_i+a^{\gamma}_i(z-z_i)^{w_i}+\cdots\,,&\quad z\to z_i\\
\gamma(z)\sim \xi_\ell+b^{\gamma}_\ell(z-\zeta_\ell)^{2}+\cdots\,,&\quad z\to\zeta_\ell\,,
\end{split}
\end{equation}
where $z_i,\zeta_\ell$ are marked points on $\Sigma$ and $a^{\alpha}_i,b^{\gamma}_\ell$ constants determined by the points $(z_i,\zeta_\ell)$ and $(x_i,\xi_\ell)$. For the moment, let us assume that $\Sigma$ is a connected surface of genus $g$ -- this turns out to capture the contributions to \eqref{eq:naive-single-m} which come from a single string worldsheet in the bulk. Requiring the map $\gamma$ to be holomorphic puts severe restrictions on the allowed moduli of the surface $\Sigma$ and marked points $z_i,\zeta_\ell$. Indeed, it can be shown that the allowed set $(\Sigma,z_i,\zeta_\ell)\in\mathcal{M}_{g,n+m}$ is finite for a given fixed genus $g$. The genus $g$ of the covering space is determined by the number of poles $N$ of $\gamma$ by the Riemann-Hurwitz relation
\begin{equation}
N=1-g+\sum_{i=1}^{n}\frac{w_i-1}{2}+\frac{m}{2}\,.
\end{equation}
The locations $\lambda_a$ of the poles are determined uniquely once a covering space is fixed, and we denote by $c^\gamma_a$ the residues of $\gamma$ at $z=\lambda_a$, i.e.
\begin{equation}
\gamma(z)\sim\frac{c^{\gamma}_a}{z-\lambda_a}\,,\quad z\to\lambda_a\,.
\end{equation}
We emphasize that the values of $c^{\gamma}_a$ are determined uniquely once a covering surface is chosen. Finally, it is convenient to introduce a constant $C^{\gamma}$ such that
\begin{equation}
\partial\gamma(z)=C^{\gamma}\frac{\prod_{i=1}^{n}(z-z_i)^{w_i-1}\prod_{\ell=1}^{m}(z-\zeta_{\ell})}{\prod_{a=1}^{N}(z-\lambda_a)^2}\,.
\end{equation}
Again, $C^{\gamma}$ is determined uniquely by the covering surface data.

The correlation function \eqref{eq:naive-single-m} is then determined purely in terms of the analytic data of the covering maps $\gamma:\Sigma\to\text{S}^2$ as well as the data of the seed CFT $\mathbb{R}_{\mathcal{Q}}\times X$. Specifically focusing on covering spaces of genus $g=0$ the correlator \eqref{eq:naive-single-m} can be written in terms of the data $(a_i^{\gamma},b_{\ell}^{\gamma},c_a^{\gamma},C^{\gamma})$ as \cite{Dei:2019iym,Hikida:2020kil,Eberhardt:2021vsx}\footnote{The factor of $|C^{\gamma}|^k$ is usually omitted in the literature, but can be derived from a careful path integral analysis of the symmetric orbifold. Alternatively, it is necessary for the correlator to obey the conformal Ward identities. See \cite{Dei:2023ivl} for a discussion on this point.}
\begin{equation}\label{eq:sym-correlators}
\begin{split}
&\sum_{\substack{\text{connected covering}\\\text{spaces }\gamma}}K^{1-\frac{n+m}{2}}|C^{\gamma}|^k\prod_{a=1}^{N}|c^{\gamma}_a|^{-k}\prod_{i=1}^{n}\left(w_i^{\frac{1}{2}-\frac{k(w_i+1)}{2}}|a_i^{\gamma}|^{-2h_i+\frac{k(w_i-1)}{2}}\right)\\
&\hspace{4cm}\times\prod_{\ell=1}^{m}\left(2^{\frac{1}{2}-\frac{3k}{2}}|b_{\ell}^{\gamma}|^{\frac{k-4}{2}}\right)\Braket{\prod_{i=1}^{n}e^{q_i\varphi}(z_i)\prod_{\ell=1}^{m}e^{\alpha\varphi}(\zeta_{\ell})}\,.
\end{split}
\end{equation}
The various factors of $2$ and $w_i$ arise from a careful analysis of the normalization of twist fields and the combinatorial factors arising in the large $K$ expansion.\footnote{We note that, crucially, equation \eqref{eq:sym-correlators} should be thought of as asymptotic in $K$. At finite $K$, the dependence on $K$ is much more complicated. This is in contrast to, say, $\text{U}(N)$ gauge theories, for which a single Feynman diagram admits a unique power of $N$, even at finite $N$.} Here, $h_i$ are the conformal weights of the states in the seed CFT, i.e. those given by equation \eqref{eq:symmetric-orbifold-conformal-weights}. The power of $K$ is $\frac{1}{2}\chi$, where $\chi$ is the Euler character of a sphere with $n+m$ punctures.

Still focusing on $g=0$, the seed CFT correlators are readily computed by Wick contractions:
\begin{equation}
\begin{split}
\Braket{\prod_{i=1}^{n}e^{-q_i\varphi}(z_i)\prod_{\ell=1}^{m}e^{-\alpha\varphi}(\zeta_{\ell})}=&\prod_{i<j}|z_i-z_j|^{-q_iq_j}\prod_{i,\ell}|z_i-\zeta_{\ell}|^{-q_i\alpha}\prod_{\ell<k}|\zeta_{\ell}-\zeta_{k}|^{-\alpha^2}\\
&\hspace{3cm}\times\delta\bigg(\sum_{i=1}^{n}q_i+m\alpha-\mathcal{Q}(1-g)\bigg),
\end{split}
\end{equation}
so that the naive perturbative answer \eqref{eq:naive-perturbative-answer} evaluates to
\begin{equation}\label{eq:naive-perturbative-expansion}
\begin{split}
K\sum_{m=0}^{\infty}&\frac{\left(-K^{-\frac{1}{2}}2^{\frac{1}{2}-\frac{3k}{2}}\mu\right)^m}{m!}\sum_{\substack{\text{connected covering}\\\text{spaces }\gamma}}\prod_{i=1}^{n}\left(K^{-\frac{1}{2}}w_i^{\frac{1}{2}-\frac{k(w_i+1)}{2}}\right)\\
&\times\int\mathrm{d}^2\xi_1\ldots\mathrm{d}^2\xi_m\,|C^{\gamma}|^k\prod_{a=1}^{N}|c^{\gamma}_a|^{-k}\prod_{i=1}^{n}|a_i^{\gamma}|^{-2h_i+\frac{k(w_i-1)}{2}}\prod_{\ell=1}^{m}|b_{\ell}^{\gamma}|^{\frac{k-4}{2}}\\
&\times\prod_{i<j}|z_i-z_j|^{-q_iq_j}\prod_{i,\ell}|z_i-\zeta_{\ell}|^{-q_i\alpha}\prod_{\ell<k}|\zeta_{\ell}-\zeta_{k}|^{-\alpha^2}\delta\bigg(\sum_{i=1}^{n}q_i+m\alpha-\mathcal{Q}(1-g)\bigg).
\end{split}
\end{equation}
Note that we have collected all of the numerical prefactors into the first line. Since $m\geq 0$ is an integer, we see that this correlation function diverges whenever the momenta $q_i$ satisfy
\begin{equation}\label{eq:naive-pole-locations}
\sum_{i=1}^{n}q_i=\mathcal{Q}(1-g)-m\alpha\,,\quad m\in\mathbb{Z}_{\geq 0}\,.
\end{equation}
As mentioned above, in the full interacting CFT, the correlators also diverge at these values of the momenta. In fact, the correlators \eqref{eq:goal-cft-correlator} are analytic functions in the momenta $q_i$ which have poles at these points. The residues of these poles are given by\footnote{The fact that residues are related to coefficients of delta functions can also be understood from the formula $\frac{1}{x-i\epsilon}=\pi\delta(x)+\mathcal{P}(\frac{1}{x})$, where $\mathcal{P}(\frac{1}{x})$ denotes the principal value. Note also that the divergence on the right-hand-side comes from the delta function since $\mathcal{P}(\frac{1}{0})=0$.}
\begin{equation}
\mathop{\mathrm{Res}}_{\sum_{i}q_i-\mathcal{Q}(1-g)=-m\alpha}\Braket{\prod_{i=1}^{n}\sigma_{w_i,q_i}(x_i)}_{\mu}=\Braket{\prod_{i=1}^{n}\sigma_{w_i,q_i}(x_i)\prod_{\ell=1}^{m}\sigma_{2,\alpha}}_{\mu=0}'\,,
\end{equation}
where the prime means the correlator without the momentum-conserving delta function inserted. Intuitively, the full interacting correlation function has all of the divergences of \eqref{eq:naive-perturbative-expansion}, but with the delta functions `smoothed out' into poles in momentum space.

In addition to the poles located at \eqref{eq:naive-pole-locations}, the reflection symmetry $q_i\to\mathcal{Q}-q_i$ means that there are poles located at all of the reflected values of the momenta. The residues of these poles will be weighted with appropriate factors of the reflection coefficients $R(h_i,\mathcal{Q}-q_i)$.

\subsection[Differences at \texorpdfstring{$k\leq 3$}{k<=3} vs \texorpdfstring{$k>3$}{3}]{\boldmath Differences at \texorpdfstring{$k\leq 3$}{k<=3} vs \texorpdfstring{$k>3$}{3}}

In the above discussion, we were simply interested in the poles of CFT correlators as analytic functions of the momenta $q_i$. These poles occur when the momenta $q_i$ satisfy the linear relation
\begin{equation}\label{eq:momentum-conservation-3.4}
\sum_{i=1}^{n}q_i+m\alpha=\mathcal{Q}\,,\quad m\in\mathbb{Z}_{\geq 0}\,,
\end{equation}
as well as all images under the reflection symmetry $q_i\to\mathcal{Q}-q_i$. However, the spectrum of the theory only includes momenta on the critical line $\text{Re}(q_i)=\mathcal{Q}/2$ in the complex plane. This implies that
\begin{equation}
\text{Re}\left(\sum_{i=1}^{n}q_i+m\alpha-\mathcal{Q}\right)=\frac{\mathcal{Q}}{2}(n-2)+\frac{m}{Q}\,.
\end{equation}
Note that if $\mathcal{Q}>0$, then there is no way to satisfy the momentum conservation \eqref{eq:momentum-conservation-3.4} if $n>2$. From the path integral, this can be seen as follows. After passing to the covering space, the integral over the zero mode $\varphi_{\infty}$ of $\varphi$ takes the form
\begin{equation}
\int^{\infty}\mathrm{d}\varphi_{\infty}\,\exp\left(-\left(\sum_{i=1}^{n}q_i+m\alpha-\mathcal{Q}(1-g)\right)\varphi_{\infty}\right)\,,
\end{equation}
where we are agnostic about the lower-bound of integration. For
\begin{equation}\label{eq:zero-mode-integral}
\text{Re}\left(\sum_{i=1}^{n}q_i+m-\mathcal{Q}(1-g)\right)=\frac{\mathcal{Q}}{2}(n+2g-2)+\frac{m}{Q}>0\,,
\end{equation}
the worldsheet integral is well-defined and there are no poles in the correlation functions. For generic correlators, this inequality is satisfied only if $\mathcal{Q}\geq 0$ or, in $\text{AdS}_3$ terms, $k\leq 3$. If $k>3$, then the above inequality is in general not satisfied and the path integral is not well-defined. In \cite{Balthazar:2021xeh}, it was argued the divergence of the linear dilaton path integral is a sign that the boundary CFT is only well-defined for $k<3$, corresponding to `sub-stringy' values of the $\text{AdS}_3$ radius.

However, as was noted in \cite{Eberhardt:2021vsx,Dei:2022pkr}, worldsheet correlators in the dual CFT are analytic in the momenta $q_i$. Thus, while the zero-mode integral formally diverges for the values of $q_i$ which live in the CFT spectrum, one can analytically continue the answer of the integral to be:
\begin{equation}
\begin{split}
&\int^{\infty}\mathrm{d}\varphi_{\infty}\,\exp\left(-\left(\sum_{i=1}^{n}q_i+m\alpha-\mathcal{Q}(1-g)\right)\varphi_{\infty}\right)\\
&\hspace{5cm}\sim\left(\sum_{i=1}^{n}q_i+m\alpha-\mathcal{Q}(1-g)\right)^{-1}\,.
\end{split}
\end{equation}
The divergences now manifest as poles in the correlation functions. Thus, while the theory naively seems to only be well-defined for $k\leq 3$, the analyticity of the correlators seems to be enough to define correlators in the regime $k>3$, at least perturbatively in $\mu$. Thus, since we are only interested in the analytic behavior of the correlators in $q_i$, we remain agnostic as to whether the CFT is defined in the regime $k>3$.

\section{Computing the string correlators}\label{sec:correlators}

In this section we will show that the worldsheet correlators in the near-boundary limit of $\text{AdS}_3$ string theory computes the naive perturbative correlation functions in the dual CFT described in the previous section. We will focus entirely on the leading contribution in the string coupling, which is dominated on the worldsheet side by genus-zero worldsheets, and on the CFT side by covering maps of genus zero, see equation \eqref{eq:sym-correlators}.

The statement of the duality (at tree-level) is that the integrated worldsheet correlators should precisely reproduce the (connected) CFT correlators. Schematically,
\begin{equation}
\begin{aligned}
g_s^{-2}\int_{\mathcal{M}_{0,n}}\Braket{\prod_{i=1}^n V_i(x_i,z_i)}=\Braket{\prod_{i=1}^n \mathcal{O}_i^{\rm CFT}(x_i)}_{\text{tree-level}}\,,
\label{mainresult1}
\end{aligned}
\end{equation}
where $\mathcal{M}_{0,n}$ is the moduli space of $n$-punctured spheres and the subscript on the CFT correlator reminds us that we are taking connected components which contribute at leading order in $1/K$. Here, $V_i$ are vertex operators on the worldsheet and $\mathcal{O}_i^{\rm CFT}$ are the corresponding states in the dual CFT. Actually, the result we want to show is slightly weaker, since we should allow for relative normalizations $\mathcal{N}_i$ of the vertex operators on both sides, as well as an overall relative normalization $C_{\text{S}^2}$ related to the definitions of the path integrals on the string theory and CFT sides:
\begin{equation}
\begin{aligned}
g_s^{-2}\int_{\mathcal{M}_{0,n}}\Braket{\prod_{i=1}^n V_i(x_i,z_i)}=\left(C_{\text{S}^2}\prod_{i=1}^{n}\mathcal{N}_i\right)\Braket{\prod_{i=1}^n \mathcal{O}_i^{\rm CFT}(x_i)}_{\text{tree-level}}\,.
\label{mainresult2}
\end{aligned}
\end{equation}

Ultimately, proving equation \eqref{mainresult2} is out of our reach, since neither the worldsheet string theory nor the dual CFT have been completely solved. However, as we have reviewed in Sections \ref{sec:ads3-strings} and \ref{sec:cft-dual}, both theories have `subsectors' which are solvable. On the string theory side, this is the limit $\Phi\to\infty$, in which strings are taken to lie near the asymptotic boundary of $\text{AdS}_3$, and for which the worldsheet sigma model becomes a free theory (with a screening operator). On the dual CFT side, the simple subsector is the limit $\varphi\to\infty$, for which we can treat the perturbative expansion \eqref{eq:naive-perturbative-expansion} seriously.

The goal of this section is to show that the correlators computed from the respective weak-coupling sectors precisely agree. The strategy is to take the near-boundary limit $\Phi\to\infty$ seriously and compute correlation functions in the resulting worldsheet theory in the path integral. Since the worldsheet theory becomes free, and the $\beta\gamma$ integral contains a large number of delta-functions, we find that this calculation can be done in a straightforward manner. As we will show below, the integrated correlator in the near-boundary theory precisely reproduces the naive perturbation series \eqref{eq:naive-perturbative-answer} in the dual CFT.

\subsection{Setting the stage}\label{sec:setting-the-stage}

Let us walk through the steps of computing the worldsheet path integral. We largely follow the exposition of \cite{Knighton:2023mhq}. We will postpone the main technical challenges to the next subsection.

Throughout this section, we focus on genus zero worldsheet correlators, but will keep the genus generic in formulae which apply more generally. At genus zero, the only worldsheet moduli are the $n-3$ cross ratios parametrizing $\mathcal{M}_{0,n}$. One can use the $\rm SL(2,\mathbb{C})$ symmetry to fix three positions of the vertex operators with the price of introducing three conformal $c$ ghost insertions at those points. Thus, a worldsheet amplitude takes the form
\begin{equation}
\begin{aligned}
\int \mathrm{d}^2z_4\cdots\mathrm{d}^2z_n \Braket{c\bar{c}V_1(x_1,z_1)c\bar{c}V_2(x_2,z_2)c\bar{c}V_3(x_3,z_3)\prod_{i=4}^{n} V_i(x_i,z_i)}\,,
\label{}
\end{aligned}
\end{equation}
where we keep the coordinates $z_1,z_2,z_3$ fixed. For simplicity, we will consider vertex operators of the form
\begin{equation}
V(x,z)=\Phi_{m,\bar{m},j}^{w}(x,z)V^{X}(z)\,,
\end{equation}
where $\Phi_{m,\bar{m},j}^{w}(x,z)$ is defined in \eqref{leadingverads3}\footnote{We do not lose any generality by assuming this since the full $\rm AdS_3$ vertex operator is given in terms of $\Phi_{m,\bar{m},j}^{w}(x,z)$ as in \eqref{fullverads3}.} and $V^X(z)$ is a vertex operator in the $X$ theory, chosen so that the full vertex operator $V_{i}$ has worldsheet conformal weight $(1,1)$. This, and the fact that the action is a sum of $\rm AdS_3$ and $X$ actions, means that we can compute worldsheet correlators by separately computing each contribution. That is, the worldsheet correlator factorizes as
\begin{equation}
|z_{12}z_{23}z_{13}|^2\int\mathrm{d}^2z_4\cdots\mathrm{d}^2z_n\Braket{\prod_{i=1}^{n}\Phi_{m_i,\bar{m}_i,j_i}^{w_i}(x_i,z_i)}\Braket{\prod_{i=1}^{n}V^X_i(z_i)}\,.
\label{eq:starting point of path integral}
\end{equation}
The prefactor $|z_{12}z_{23}z_{13}|^2$ comes from the Wick contraction of the $c$ ghosts. For the rest of this section, we will implicitly consider only correlators which are trivial with respect to the compact CFT $X$. This is done for notational convenience, and re-introducing the $X$-dependence of correlators is straightforward.

Let us now consider the contribution from $\rm AdS_3$. Since the expression \eqref{fullverads3} is the leading order in the large $\Phi$ expansion, we should therefore keep only the leading term in the action \eqref{fullaction} at large $\Phi$. This amounts to dropping the interaction term $\nu^{-1}\beta\bar{\beta}e^{-Q\Phi}$ and working with the resulting action
\begin{equation}
S_{\rm AdS_3}=\frac{1}{2\pi}\int\left(\frac{1}{2}\partial\Phi\,\overline{\partial}\Phi+\beta\overline{\partial}\gamma+\bar{\beta}\partial\bar{\gamma}-\frac{Q}{4}R\Phi-pD\right)\,.
\label{freeaction}
\end{equation}
With this action, let us now massage the integrand of \eqref{eq:starting point of path integral} into a more suggestive form. By expanding in powers of the screening operator $D$, the $\rm AdS_3$ correlator in \eqref{eq:starting point of path integral} becomes (at order $N$)
\begin{equation}
\frac{p^N}{N!}\int\prod_{a=1}^N\mathrm{d}^2\lambda_a\Braket{\prod_{a=1}^{N}D(\lambda_a)\prod_{i=1}^n \Phi_{m_i,\bar{m}_i,j_i}^{w_i}(x_i;z_i)}_{\text{free}}\,,
\label{orderNamp}
\end{equation}
where the subscript \emph{free} on the correlator reminds us that we are calculating the correlator in the free theory, i.e. with the action \eqref{freeaction} with $p=0$.

This string correlator further factorizes into three components: namely, that of the scalar $\Phi$, that of the $\beta\gamma$ system of the Wakimoto representation and that of the compact CFT $X$. The compact CFT $X$ simply comes along for the ride and plays no central role until we discuss the dual CFT. The scalar correlator in the near-boundary (large $\Phi$) limit takes the form
\begin{equation}
\Braket{\prod_{a=1}^{N}e^{-2\Phi/Q}(\lambda_a)\prod_{i=1}^{n}e^{-(Qj_i-w_i/Q)\Phi}(z_i)}_{\text{free}}\,,
\label{phicorr}
\end{equation}
while the $\beta\gamma$ correlator is better written in the path integral language:\footnote{From now on, we set $m_i=\bar{m}_i$. The generalization to $m_i\neq \bar{m}_i$ is immediate but clutters notation significantly.}
\begin{equation}
\int\mathcal{D}(\beta,\gamma)\,e^{-S[\beta,\gamma]}\prod_{a=1}^{N}\left|\oint_{\lambda_a}\gamma\right|^{-2(k-1)}\delta^{(2)}(\beta(\lambda_a))\prod_{i=1}^{n}\left|\frac{\partial^{w_i}\gamma(z_i)}{w_i!}\right|^{-2m_i-2j_i}\delta^{(2)}_{w_i}(\gamma(z_i)-x_i)\,.
\label{betagammapart}
\end{equation}

Due to the symmetry $\Phi\to\Phi+\Phi_0$, the correlator \eqref{phicorr} vanishes unless the charge conservation condition
\begin{equation}\label{eq:phi-charge-cons}
\frac{2N}{Q}+\sum_{i=1}^{n}\left(Qj_i-\frac{w_i}{Q}\right)=Q(1-g)\,,
\end{equation}
is satisfied. The right-hand-side comes from the background charge in the action \eqref{freeaction}. This conservation law determines the allowed value of $N$ in terms of the quantum numbers of the vertex operators:
\begin{equation}
N=1-g+\sum_{i=1}^{n}\frac{w_i-1}{2}-\frac{Q^2}{2}\left(\sum_{i=1}^{n}j_i-\frac{k}{2}(n+2g-2)+(n+3g-3)\right)\,.
\label{counting1}
\end{equation}
Since $N,g,w_i$ are integers, we immediately have
\begin{equation}
\sum_{i=1}^{n}j_i-\frac{k}{2}(n+2g-2)+(n+3g-3)=-\frac{m}{Q^2}\,,
\label{jconstraint}
\end{equation}
where $m\in\mathbb{Z}$. As we will see below, the worldsheet path integral vanishes unless $m\geq 0$.

Let us now turn to the $\beta\gamma$ path integral \eqref{betagammapart}. Following \cite{Knighton:2023mhq}, we can formally consider the Fourier transform
\begin{equation}
\delta^{(2)}(\beta(\lambda_a))=\int\mathrm{d}^2\xi_a\,e^{i\beta(\lambda_a)\xi_a+i\bar{\beta}(\bar{\lambda}_a)\bar{\xi}_a} 
\label{formalid}
\end{equation}
and thus we can think of the insertion of the delta functions $\delta^{(2)}(\beta(\lambda_a))$ in the path integral \eqref{betagammapart} as a modification of the action of the $\beta\gamma$ system. The modified action reads
\begin{equation}
S_{\text{new}}=\frac{1}{2\pi} \int_\Sigma\left[\beta\left( \bar\partial\gamma-2\pi i\sum_a\xi_a\delta^{(2)}(z,\lambda_a) \right)+\bar{\beta}\left(\partial\bar{\gamma}-2\pi i\sum_{a}\bar{\xi}_a\delta^{(2)}(z,\lambda_a)\right)\right].
\end{equation}
Integrating out $\beta,\bar{\beta}$ in \eqref{betagammapart} introduces a Dirac functional
\begin{equation}
\int \mathcal{D}\beta\exp(-S_{\text{new}})=\delta^{(2)}\bigg( \bar\partial\gamma-2\pi i\sum_a\xi_a\delta^{(2)}(z-\lambda_a)\bigg)
\end{equation}
which imposes that $\gamma$ is meromorphic with $N$ simple poles at $\lambda_a$ with residue $\xi_a$ as well as analogous conditions for $\bar\gamma$. However, $\lambda_a$ and $\xi_a$ are not fixed but are integrated over as can be seen from \eqref{formalid} and \eqref{orderNamp}. Hence, the remaining path integral over $\gamma$ is an integral over the space of meromorphic functions with $N$ simple poles and we denote this space by $\mathcal{F}_N$. This space has (complex) dimension
\begin{equation}
\text{dim}(\mathcal{F}_N)=2N+1-g\,,
\end{equation}
which can be derived from the Riemann-Roch theorem.\footnote{See for example footnote 20 of \cite{Dei:2023ivl}.} Schematically, we can thus think of the string path integral as
\begin{equation}\label{eq:integral-dimension-reduction}
|z_{12}z_{23}z_{13}|^2\int_{\mathcal{F}_N}\braket{\beta\gamma}\braket{\Phi}\braket{X}\,,
\end{equation}
where $\braket{\beta\gamma}$ is the integrand coming from the Wakimoto field $\gamma$, $\braket{\Phi}$ is the correlator \eqref{phicorr}, and $\braket{X}$ is the compact CFT correlator. As mentioned above, we will set $\braket{X}=1$ for simplicity, although it is trivial to generalize the discussion to nontrivial correlation functions in the compact CFT.

Now, the delta functions $\delta^{(2)}_{w_i}(\gamma(z_i)-x_i)$ in the path integral \eqref{betagammapart} impose $\sum_i w_i$ additional constraints on the space of functions $\gamma$. These constraints require that $\gamma$ obeys
\begin{equation}
\gamma(z)=x_i+\mathcal{O}((z-z_i)^{w_i}),\text{ as } z\to z_i
\end{equation}
for all $i=1,\ldots,n$. The space of functions that satisfy these conditions now has dimension
\begin{equation}
\text{dim}(\mathcal{F}_N)-\sum_iw_i=-Q^2\left( \sum_ij_i-\frac{k}{2}(n+2g-2)+(n+3g-3) \right)-(n+3g-3)\,,
\end{equation}
where we have used the charge conservation \eqref{counting1} to rewrite $N$ in terms of the $\text{SL}(2,\mathbb{R})$ spins $j_i$. In the end, we also want to integrate not only over the functions $\gamma$ but also over the worldsheet moduli, as in equation \eqref{mainresult1}. This means that, after computing the $\Phi$ correlator \eqref{phicorr}, the worldsheet path integral reduces to an integral of dimension
\begin{equation}
\text{dim}(\mathcal{F}_N)-\sum_{i=1}^{n}w_i+\text{dim}(\mathcal{M}_{g,n})=2N+2g-2-\sum_{i=1}^{n}(w_i-1)
\end{equation}
In order for the path integral to make any sense, this dimension needs to be a non-negative integer, which we call $m\in\mathbb{Z}_{\geq 0}$. Using the charge conservation \eqref{counting1}, we can write $m$ as a function of the $\text{SL}(2,\mathbb{R})$ spins $j_i$:
\begin{equation}\label{eq:m-definition-string-side}
m=-Q^2\left( \sum_ij_i-\frac{k}{2}(n+2g-2)+(n+3g-3) \right)
\end{equation}

The number $m$ has a simple interpretation from the point of view of integration over maps $\gamma$. Since $\gamma$ has precisely
\begin{equation}
N=1-g+\sum_{i=1}^{n}\frac{w_i-1}{2}+\frac{m}{2}\,,
\end{equation}
poles, the Riemann-Hurwitz formula tells us that $m$ counts the number of `extra' branch points of $\gamma$, i.e. the number of points $\zeta_\ell$ on the worldsheet where $\gamma$ takes the local form
\begin{equation}
\gamma(\zeta_\ell)\sim\xi_\ell+\mathcal{O}((z-\zeta_\ell)^2)\,,\quad z\to\zeta_\ell\,.    
\end{equation}
Hence, the worldsheet map $\gamma$ has precisely the behavior as the covering maps which calculate the dual CFT correlators at $m^{\text{th}}$ order in perturbation theory (see Section \ref{subsec:cft-correlators}).

\subsection{Evaluating the path integral}

Now that we have set the stage, we can evaluate the path integral \eqref{orderNamp}. This calculation has three steps:
\begin{enumerate}

    \item For a fixed value of $N$, evaluate the $\Phi$ correlator \eqref{phicorr}.

    \item Evaluate the $\beta\gamma$ path integral \eqref{betagammapart}.

    \item Massage the end result to take the form \eqref{eq:naive-perturbative-expansion} of the dual CFT correlators.
    
\end{enumerate}
As we outlined above, the result of Step 1 will be a set of Wick contractions, along with a delta function enforcing the charge conservation \eqref{eq:phi-charge-cons}, which picks out the specific value \eqref{counting1} for the number of operators $D$ inserted into the path integral. For Step 2, we then integrate out $\beta,\bar{\beta}$ and we are left with an integral over the set of meromorphic functions $\gamma$ with poles at $\lambda_a$. The delta functions in the path integral reduce this to a finite-dimensional integral of dimension $m$, defined in \eqref{eq:m-definition-string-side}. Upon calculating the appropriate Jacobians from the delta function, multiplying by $|z_{12}z_{23}z_{13}|^2$, and integrating over the worldsheet moduli $z_4,\ldots,z_n$ then gives the full string correlator. Finally, for Step 3, we will find that the Wick contractions of the $\Phi$ correlator can be repackaged nicely into a simple function of the covering map data which enters the symmetric orbifold calculation \eqref{eq:naive-perturbative-expansion}.

\subsubsection*{Step 1:}

The scalar field $\Phi$ enters the string correlator through the contribution \eqref{phicorr}. As $\Phi$ is a free scalar, this correlation function can be computed via Wick contractions:
\begin{equation}
\begin{aligned}
&\Braket{\prod_{a=1}^{N}e^{-2\Phi/Q}(\lambda_a)\prod_{i=1}^{n}e^{(w_i/Q-Qj_i)\Phi}(z_i)}_{\text{free}}=\prod_{a<b}|\lambda_a-\lambda_b|^{-8/Q^2}\prod_{i,a}|z_i-\lambda_a|^{4(w_i/Q^2-j_i)}\\
&\hspace{2.5cm}\times\prod_{i<j}|z_i-z_j|^{-2(w_i/Q-j_iQ)(w_j/Q-j_jQ)}\delta\bigg(\frac{2N}{Q}+\sum_{i=1}^{n}\left(Qj_i-\frac{w_i}{Q}\right)-Q\bigg).
\label{phicorrwick}
\end{aligned}
\end{equation}
The delta-function comes from the integral over the zero mode of $\Phi$, and ensures that the operators in \eqref{phicorr} satisfy the anomalous charge conservation \eqref{counting1}.

\subsubsection*{Step 2:}

Let us now treat the $\beta\gamma$ path integral. In order to make any sense of the integral \eqref{eq:integral-dimension-reduction}, we need to specify a measure on the space of meromorphic functions. Here, we only work for genus $g=0$ worldsheets, but we expect that it is possible to generalize it to higher genus.

As we have seen from the previous subsection, integrating out the $\beta$ field restricts $\gamma$ to lie in the space $\mathcal{F}_N$ of meromorphic functions with $N$ poles. Thus, by picking a suitable basis for this class of functions, we can deduce the path integral measure. After integrating out $\beta$, the path integral looks like
\begin{equation}
p^N\int_{\mathcal{A}_N}\mathcal{D}\gamma\prod_{a=1}^{N}\left|\oint_{\lambda_a}\gamma\right|^{-2(k-1)}\prod_{i=1}^{n}\left|\frac{\partial^{w_i}\gamma(z_i)}{w_i!}\right|^{-2m_i-2j_i}\delta^{(2)}_{w_i}(\gamma(z_i)-x_i)\,,
\label{gammapathint}
\end{equation}
where $\mathcal{A}_N$ denotes the space of meromorphic functions with $N$ simple poles at the \textit{fixed} points $\lambda_a$, but with arbitrary residues. These points are fixed since we have not yet integrated over $\lambda_a$.

Focusing on $g=0$, the most general form of such a function is
\begin{equation}
\gamma(z)=b+\sum_{a=1}^N\frac{c_a^{\gamma}}{z-\lambda_a}\,.
\label{gammaexpansion}
\end{equation}
Note that there can be no polynomial in $z$ since it would imply that $z=\infty$ has a pole singularity. Hence, the space $\mathcal{A}_N$ is parametrized by $b,c_a^{\gamma}$ and thus it is natural to define the path integral measure\footnote{One may argue that there should be some nontrivial Jacobian at this step. We refer to Appendix \ref{sec:deriving path integral measure} for a more detailed argument why there should not be a nontrivial Jacobian associated to this change of variable.}
\begin{equation}
\int_{\mathcal{A}_N}\mathcal{D}\gamma=\int \mathrm{d}^2b\prod_{a=1}^N\mathrm{d}^2c_a^{\gamma}\,.
\label{measuregamma}
\end{equation}
With this, we can now write eq. \eqref{gammapathint} as
\begin{equation}
p^N\int \mathrm{d}^2b\,\mathrm{d}^{2N}c_a^{\gamma}\prod_{a=1}^{N}\left|c_a^{\gamma}\right|^{-2(k-1)}\prod_{i=1}^{n}\left|\frac{\partial^{w_i}\gamma(z_i)}{w_i!}\right|^{-2m_i-2j_i}\delta^{(2)}_{w_i}(\gamma(z_i)-x_i)\,,
\label{prefinalgammapathint}
\end{equation}
with $\gamma(z)$ given by \eqref{gammaexpansion}. If we now include the integration over $\lambda_a$ from \eqref{orderNamp}, we obtain
\begin{equation}
\frac{p^N}{N!}\int \mathrm{d}^2b\,\mathrm{d}^{2N}\lambda_a\,\mathrm{d}^{2N}c_a^{\gamma}\,\prod_{a=1}^{N}\left|c_a^{\gamma}\right|^{-2(k-1)}\prod_{i=1}^{n}\left|\frac{\partial^{w_i}\gamma(z_i)}{w_i!}\right|^{-2m_i-2j_i}\delta^{(2)}_{w_i}(\gamma(z_i)-x_i)\,.
\label{finalgammapathint}
\end{equation}
The worldsheet amplitude \eqref{eq:starting point of path integral} now schematically takes the form
\begin{equation}
\begin{aligned}
\frac{p^N}{N!}\int \prod_{j=4}^n\mathrm{d}^2z_j\,\mathrm{d}^2b\prod_{a=1}^N\mathrm{d}^2\lambda_a\,\mathrm{d}^2c_a^{\gamma}\braket{\Phi}\braket{\beta\gamma}|z_{12}z_{23}z_{13}|^2\,,
\label{totalwsamp}
\end{aligned}
\end{equation}
where $\braket{\Phi}$ is given by \eqref{phicorrwick}, $\braket{X}$ is the correlator in the compact direction and $\braket{\beta\gamma}$ is the integrand of \eqref{finalgammapathint}. The integral is taken over the $2N+n-2$ parameters $z_j,b,\lambda_a,c_a^{\gamma}$, whereas the integrand is a delta function of the form
\begin{equation}\label{eq:collected-delta-functions}
\prod_{i=1}^{n}\prod_{\ell_i=0}^{w_i-1}\delta^{(2)}\left(\frac{\partial^{\ell_i}(\gamma(z_i)-x_i)}{\ell_i!}\right)\,.
\end{equation}
Thus, the delta functions eliminate $\sum_{i}w_i=2N+n-2-m$ of the $2N+n-2$ integrals, and we are left with an $m$-dimensional integral. As mentioned in Section \ref{sec:setting-the-stage}, $m$ is precisely the number of `extra' branch points of the covering map $\gamma$. Thus, a natural set of variables to integrate over are the arguments of the delta functions \eqref{eq:collected-delta-functions} and the values $\gamma(\zeta_\ell)$ of $\gamma$ at the $m$ branch points $\zeta_\ell$. That is, we consider the change of variables
\begin{equation}\label{eq:moduli-space-coordinate-change}
(z_j,b,\lambda_a,c_a^{\gamma})\to\left(\frac{\partial^{\ell_i}(\gamma(z_i)-x_i)}{\ell_i!},\gamma(\zeta_\ell)\right)
\end{equation}
where $j=4,\ldots,n$, $a=1,\ldots,N$, $i=1,\ldots,n$, $\ell_i=0,\ldots,w_i-1$ and $l=1,\ldots,m$. Under this change of variables, the path integral \eqref{totalwsamp} becomes\footnote{The factor of $1/m!$ comes from the fact that the coordinates on the RHS of \eqref{eq:moduli-space-coordinate-change} cover the moduli space in question $m!$ times.}
\begin{equation}
\begin{aligned}
&p^N\int \prod_{j=4}^n\mathrm{d}^2z_j\,\mathrm{d}^2b\prod_{a=1}^N\mathrm{d}^2\lambda_a\,\mathrm{d}^2c_a^{\gamma}\braket{\Phi}\braket{\beta\gamma}|z_{12}z_{23}z_{13}|^2\\
&\hspace{1cm}=\frac{p^N}{m!}\int\mathrm{d}^{2m}\gamma(\zeta_\ell)\,\mathcal{J}^{-1}\prod_{a=1}^N|c_a|^{-2(k-1)}\prod_{i=1}^n|a^\gamma_i|^{-2m_i-2j_i}\braket{\Phi}|z_{12}z_{23}z_{13}|^2\,,
\label{}
\end{aligned}
\end{equation}
where we have defined $a^\gamma_i:=\partial^{w_i}\gamma(z_i)/w_i!$ and $\mathcal{J}$ is the Jacobian of the change of variables. This Jacobian can be found by a straightforward, but tedious, calculation and we spell out the details in Appendix \ref{betagammaappen}. The end result is that $\mathcal{J}$ is given by (see \eqref{mostgenJ})
\begin{equation}\label{eq:jacobian-main-text}
\mathcal{J}=|z_{12}z_{13}z_{23}|^2|C^\gamma|^{-4}\prod_{i=1}^n|A^\gamma_i|^{w_i+1} 
\prod_{\ell=1}^m|B^\gamma_\ell|\,,
\end{equation}
with $A^\gamma_{i}$ and $B^{\gamma}_{\ell}$ are defined below as in \eqref{relationaGamma}. Defining $\xi_\ell:=\gamma(\zeta_\ell)$, we can rewrite the worldsheet amplitude as
\begin{equation}
\begin{split}
&\frac{p^N}{m!}\bigg(\prod_{i=1}^n\frac{1}{w_i^{w_i+1}2^{m}}\bigg)\int \mathrm{d}^{2m}\xi_{\ell}\,|C^\gamma|^4\prod_{a=1}^N|c_a^{\gamma}|^{-2(k-1)}\prod_{i=1}^n|a^\gamma_i|^{-2(m_i+j_i)-(w_i+1)}\prod_{\ell=1}^m|b^\gamma_\ell|^{-1}\braket{\Phi}\,.
\label{wsalmostfinished}
\end{split}
\end{equation}
Here, $C^\gamma$ is a constant defined by
\begin{equation}
\partial\gamma(z)=C^\gamma\frac{\prod_{i=1}^n(z-z_i)^{w_i-1}\prod_{l=1}^m(z-\zeta_l)}{\prod_{a=1}^N(z-\lambda_a)^2}\,,
\label{weight1dgamma}
\end{equation}
and $b_\ell^{\gamma}$ is the first nontrivial Taylor coefficient of $\gamma$ around the point $\zeta_\ell$, i.e.
\begin{equation}
\gamma(z)\sim\xi_\ell+b_\ell^{\gamma}(z-\zeta_\ell)^2\,.
\end{equation}
Using \eqref{weight1dgamma}, we also have
\begin{equation}
\begin{aligned}
A^\gamma_i:=w_ia^\gamma_i=&C^\gamma\frac{\prod_{k\neq i}^n(z_i-z_k)^{w_k-1}\prod_{l=1}^m(z_i-\zeta_l)}{\prod_{a=1}^N(z_i-\lambda_a)^2}\\
c_a^{\gamma}=&-C^\gamma\frac{\prod_{i=1}^n(\lambda_a-z_i)^{w_i-1}\prod_{l=1}^m(\lambda_a-\zeta_l)}{\prod_{b\neq a}^N(\lambda_a-\lambda_b)^2}\\
B^\gamma_l:=2b^\gamma_l=&C^\gamma\frac{\prod_{i=1}^n(\zeta_l-z_i)^{w_i-1}\prod_{p\neq l}^m(\zeta_l-\zeta_p)}{\prod_{a=1}^N(\zeta_l-\lambda_a)^2}\,,
\label{relationaGamma}
\end{aligned}
\end{equation}
where we have defined $A^{\gamma}_i,B^{\gamma}_\ell$ in terms of factors of $2$ and $w_i$ and $a_i^{\gamma},b^{\gamma}_\ell$. Note that these are precisely the covering map data used to define symmetric orbifold correlators as in Section \ref{sec:cft-dual}.

\subsubsection*{Step 3:}

We have now computed the $\beta\gamma$ path integral in terms of the covering map data $a_i^{\gamma},b_{\ell}^{\gamma},c_a^{\gamma},C^{\gamma}$. The final step is to include the $\Phi$ correlator \eqref{phicorrwick} to obtain the full string correlator.

First, note that we can replace the various factors of $|z_i-z_j|$, $|z_i-\lambda_a|$, and $|\lambda_a-\lambda_b|$ in \eqref{phicorrwick} with the covering map data. Indeed, using \eqref{relationaGamma}, we can rewrite the correlator \eqref{phicorrwick} as
\begin{equation}
\begin{aligned}
\Braket{\Phi}=&|C^\gamma|^{\frac{2}{Q^2}-2}\prod_a|c_a^{\gamma}|^{\frac{2}{Q^2}}\prod_\ell|A_\ell^\gamma|^{\frac{1}{Q^2}}\prod_i|A^\gamma_i|^{-(\frac{w_i+1}{Q^2}-2j_i)}\\
&\times\prod_{\ell<p}|\zeta_\ell-\zeta_p|^{-\frac{2}{Q^2}}
\prod_{i,\ell}|z_i-\zeta_\ell|^{-2j_i+\frac{2}{Q^2}}\prod_{i<k}|z_i-z_k|^{-2(Qj_i-\frac{1}{Q})(Qj_k-\frac{1}{Q})}\\
&\hspace{5.5cm}\times\delta\bigg(\frac{2N}{Q}+\sum_{i=1}^{n}\left(Qj_i-\frac{w_i}{Q}\right)-Q\bigg).
\label{}
\end{aligned}
\end{equation}
Inserting the $\Phi$ correlator into \eqref{wsalmostfinished} (and removing the delta function), we finally arrive at the expression

\begin{tcolorbox}
\begin{equation}
\begin{gathered}
\sum_{\substack{\text{connected covering}\\\text{maps }\gamma}}p^N\left( \prod_{i=1}^nw_i^{-\frac{k(w_i+1)}{2}+2j_i}2^{\frac{(k-4)m}{2}} \right)\int \frac{\mathrm{d}^{2m}\xi_{\ell}}{m!}\sum_{\gamma}|C^{\gamma}|^k\prod_{a=1}^{N}|c^{\gamma}_a|^{-k}\\
\prod_{i=1}^{n}|a_i^{\gamma}|^{-2h_i+\frac{k(w_i-1)}{2}}\prod_{\ell=1}^{m}|a_\ell^{\gamma}|^{\frac{k-4}{2}}\prod_{i<j}|z_i-z_j|^{-2q_iq_j}\prod_{i,\ell}|z_i-\zeta_{\ell}|^{-2q_i\alpha}\prod_{\ell<k}|\zeta_\ell-\zeta_k|^{-2\alpha^2}.
\label{wsfinished}
\end{gathered}
\end{equation}
\end{tcolorbox}

\noindent Here, $h_i$ is defined in \eqref{hhbar} and
\begin{equation}
q_i:=Qj_i-\frac{1}{Q}\,,\quad \alpha:=\frac{1}{Q}\,.
\end{equation}
Note that in \eqref{wsfinished}, we have also included the possibility that the space of covering maps is not connected, and so we sum over non-continuously related maps. Aside from the overall normalization (which we will discuss below), this expression has precisely the form of \eqref{eq:naive-perturbative-expansion} with $c=6k$.

\subsection{Matching the normalization}

To obtain a precise agreement between the CFT answer \eqref{eq:naive-perturbative-expansion} and the worldsheet answer \eqref{wsfinished}, we need to introduce various normalization constants. Specifically, we need to determine the constants $\mathcal{N}(j_i,w_i)$ and $C_{\text{S}^2}$ on the right-hand-side of equation \eqref{mainresult2}, as well as a dictionary relating the parameters $(\nu,p)$ of the worldsheet CFT to the parameter $\mu$ of the boundary CFT. It is straightforward to verify that the appropriate choice of constants is
\begin{equation}
\begin{gathered}
\mathcal{N}(j,w)=K^{1/2}\nu^{\frac{(2-k)(w-1)}{2}}w^{2j-\frac{1}{2}}\,,\quad\mu=-K^{1/2}2^{2k-\frac{5}{2}}\nu^{\frac{2-k}{2}}\,,\\
p=\nu^{2-k}\,,\quad C_{\text{S}^2}=\nu^{2-k}\,,\quad g_s=\frac{1}{\sqrt{K}}\,.
\end{gathered}
\end{equation}
With these normalizations, we arrive at a precise match between the string correlators \eqref{wsfinished} and the CFT correlators \eqref{eq:naive-perturbative-expansion}.\footnote{The various normalizations differ from those in \cite{Dei:2022pkr}, see their equation (3.2). We suspect that this difference is due to a difference in conventions for the $\text{SL}(2,\mathbb{R})$ WZW model.}

\section{Discussion}\label{sec:discussion}

In the previous section, we calculated worldsheet correlation functions of spectrally-flowed vertex operators in $\text{AdS}_3\times X$ in the limit that the strings are localized near the boundary. This region in the bulk is dual to the regime of the dual CFT where correlators can be computed using naive perturbation theory. The analysis of Section \ref{sec:correlators} showed that the string theory correlators in the near-boundary limit indeed reproduce the corresponding perturbative correlators of the dual CFT. We emphasize that this result holds for 1) all spectral flows/twists $w_i$, 2) all numbers $n$ of vertex operator insertions, and 3) at all orders in conformal perturbation theory in $\mu$. This effectively proves the proposed near-boundary/weak-coupling duality \eqref{eq:string-duality-weak} and constitutes extremely strong evidence for the full $\text{AdS}_3/\text{CFT}_2$ duality \eqref{eq:string-duality}.

Let us recap how the worldsheet calculation went. First, we had to integrate out the worldsheet Wakimoto scalar $\Phi$, which introduced various analytic factors in the form of Wick contractions. We then re-wrote these Wick contractions with the help of certain analytic data of the covering map $\gamma$. Upon computing the path integral for the $\beta\gamma$ system, we found that the worldsheet path integral localizes to covering maps with certain branching structures (precisely those appearing in the symmetric orbifold calculations). Upon computing the Jacobian in the $\beta\gamma$ path integral and combining it with the correlator of the scalar $\Phi$, we miraculously found a formula which matches the dual CFT correlator precisely. While this computation certainly \textit{worked}, it is conceivable that there is a much more straightforward way to get the answer. We will now briefly speculate how this might go. 

The worldsheet correlation functions of the $\Phi$ scalar take the form
\begin{equation}
\Braket{\prod_{a=1}^{N}e^{-2\Phi/Q}(\lambda_a)\prod_{i=1}^{n}e^{(w_i/Q-Qj_i)\Phi}(z_i)}\,,
\end{equation}
whereas the correlators of the dual linear dilaton $\varphi$ take the form (after passing to the covering space)
\begin{equation}
\Braket{\prod_{i=1}^{n}e^{(1/Q-Qj_i)\varphi}(z_i)\prod_{\ell=1}^{m}e^{-\varphi/Q}(\zeta_{\ell})}\,.
\end{equation}
By analyzing the OPE structure of the currents $\partial\Phi,\partial\varphi$, we see that within correlation functions we can make the replacements
\begin{equation}
\begin{split}
\partial\Phi(z)&=\sum_{a=1}^{N}\frac{2/Q}{z-\lambda_a}+\sum_{i=1}^{n}\frac{Qj_i-w_i/Q}{z-z_i}\,,\\
\partial\varphi(z)&=\sum_{i=1}^{n}\frac{Qj_i-1/Q}{z-z_i}+\sum_{\ell=1}^{m}\frac{1/Q}{z-\zeta_\ell}\,.
\end{split}
\end{equation}
Taking the difference yields
\begin{equation}
\partial\Phi(z)-\partial\varphi(z)=\frac{1}{Q}\left(\sum_{a=1}^{N}\frac{2}{z-\lambda_a}-\sum_{i=1}^{n}\frac{(w_i-1)}{z-z_i}-\sum_{\ell=1}^{m}\frac{1}{z-\zeta_\ell}\right)\,.
\end{equation}
Defining $\gamma(z)$ through \eqref{weight1dgamma}, we have
\begin{equation}
\partial\Phi(z)=\partial\varphi(z)-\frac{1}{Q}\partial\log\partial\gamma\,.
\end{equation}
The same analysis for the right-moving sector gives the relation
\begin{equation}\label{eq:discussion-phi-sub}
\Phi=\varphi-\frac{1}{Q}\Phi_{\gamma}\,,
\end{equation}
where we have defined 
\begin{equation}
\Phi_{\gamma}=\log|\partial\gamma|^2\,.
\end{equation}
We note that the field redefinition \eqref{eq:discussion-phi-sub} is similar to the shift of the Wakimoto scalar performed in \cite{Hikida:2007tq,Hikida:2008pe,Hikida:2020kil,Hikida:2023jyc}. It is also similar to the anomalous transformation derived in the Appendix C of \cite{Eberhardt:2019ywk}.

The reason that the shift \eqref{eq:discussion-phi-sub} is useful is that $\text{Sym}(\mathbb{R}_{\varphi}\times X)$ correlators are determined entirely by the linear dilaton field $\varphi$ and the scalar $\Phi_{\gamma}$ in a straightforward way \cite{Lunin:2000yv,Lunin:2001ne}:\footnote{We are assuming all vertex operators are in the vacuum of $X$.}
\begin{equation}\label{eq:discussion-symm-correlators}
\Braket{\prod_{i=1}^{n}\sigma_{w_i,q_i}(x_i)\prod_{\ell=1}^{m}\sigma_{2,\alpha}}=\sum_{\substack{\text{connected covering}\\\text{maps }\gamma}}e^{-S_L[\Phi_{\gamma}]}\Braket{\prod_{i=1}^{n}e^{-q_i\varphi}(z_i)\prod_{\ell=1}^{m}e^{-\alpha\varphi}(\zeta_\ell)}\,,
\end{equation}
where the linear dilaton correlator on the right-hand-side is computed on the covering space and $S_L$ is the Liouville action
\begin{equation}
S_L[\Phi_{\gamma}]=\frac{c}{48\pi}\int\mathrm{d}^2z\sqrt{g}\left(2\partial\Phi_{\gamma}\overline{\partial}\Phi_{\gamma}+R\Phi_{\gamma}\right)\,.
\end{equation}
One might therefore hope that the substitution \eqref{eq:discussion-phi-sub} would allow one to write the string path integral manifestly in the form \eqref{eq:discussion-symm-correlators} without the need of systematically computing Wick contractions and determinants of the $\beta\gamma$ system. This would have the advantage of 1) laying bare the precise mechanism of how the string path integral is repackaged into a symmetric orbifold correlation function and 2) working for worldsheets of generic genus, i.e. at all orders in the $g_s$ expansion. This would effectively amount to a direct derivation of the $\text{AdS}_3/\text{CFT}_2$ correspondence \eqref{eq:string-duality-weak} in the perturbative regime. As we still do not fully understand the details of how such a derivation would go, we leave this to future work.

\subsection*{Future directions}

We will now discuss several future directions that this work could take, as well as problems which we believe are now approachable given the progress made in this paper.

\paragraph{Worldsheet perturbations and Coulomb gas integrals:} In this paper we matched the near-boundary worldsheet path integral with the naive perturbative expansion of the dual CFT. On the worldsheet, considering the near-boundary limit effectively amounts to ignoring the interaction term $\nu^{-1}\beta\bar{\beta}e^{-Q\Phi}$ in the AdS Lagrangian. The perturbative expansion of the dual CFT is then reproduced by the perturbative expansion in the screening operator $D$ on the worldsheet. A natural extension of the results of this paper would be to consider the effect of perturbatively introducing the interaction term in the $\text{AdS}_3$ sigma model. From the worldsheet perspective, perturbing in $\beta\bar{\beta}e^{-Q\Phi}$ would introduce a 2-dimensional lattice of poles where
\begin{equation}\label{eq:pole-locations-conclusions}
\frac{2N}{Q}+MQ+\sum_{i=1}^{n}\left(Qj_i-\frac{w_i}{Q}\right)=Q(1-g)\,,\quad M,N\in\mathbb{Z}_{\geq 0}\,.
\end{equation}
Of this set of poles, only the set $M=0$ is understood from the dual CFT perspective. Under the assumption that the full duality \eqref{eq:string-duality} holds, it would be interesting to understand the roles that the poles $M>0$ play in the dual CFT. The situation is likely to be similar to Liouville theory, where the duality $b\to 1/b$ implies the existence of a two-dimensional lattice of poles which are not visible in naive perturbation theory. It would thus be fruitful to explore whether a technique analogous to the coulomb gas formalism for Liouville theory \cite{Goulian:1990qr} exists for the CFT dual of $\text{AdS}_3$ string theory. If such a formalism does exist, then it would be in principle possible to match the residues at all of the poles in \eqref{eq:pole-locations-conclusions}, and not just those with $M=0$.

\paragraph{Worldsheet perturbation and the tensionless limit}
It was speculated in \cite{McStay:2023thk,Dei:2023ivl} that the interaction term $\beta\bar\beta e^{-Q\Phi}$ does not give any contribution in the tensionless limit. In other words, the near-boundary limit becomes exact in the tensionless limit. However, the precise mechanism that explains why this is the case is still lacking in the literature. Indeed, by treating the interaction term perturbatively and restricting ourselves to the tensionless limit ($k=3$), it can be shown that the interaction gives no contribution in generic string amplitude. This is currently under investigation \cite{Sriprachyakul:2024abc}.

\paragraph{\boldmath Solving the $\text{SL}(2,\mathbb{R})$ WZW model:}

String theory on $\text{AdS}_3$ is based on the $\text{SL}(2,\mathbb{R})$ WZW model (or $\text{SU}(2,\mathbb{C})/\text{SU}(2)$ in Euclidean signature). As we have seen in this paper, a careful treatment of the free field realization allows one to exactly calculate spectrally-flowed correlation functions in the near-boundary limit. As mentioned above, a Coulomb gas-type formalism including the interaction term $\nu^{-1}\beta\bar{\beta}e^{-Q\Phi}$ could allow us to identify all of the poles in the correlators of the $\text{SL}(2,\mathbb{R})$ WZW model. Another, more ambitious, goal would be to use the resulting Coulomb gas formalism to completely solve the $x$-basis correlators of the $\text{SL}(2,\mathbb{R})$ WZW model, similar to the solution of Liouville theory \cite{Dorn:1994xn,Zamolodchikov:1995aa}.

\paragraph{\boldmath Superstrings on $\text{AdS}_3\times X$:}

Another natural direction to persue is to supersymmetrize our results and check the validity of the superstring proposal by Eberhardt \cite{Eberhardt:2021vsx}. In fact, one can already see from our result that a naive guess for the CFT dual to superstring in ${\rm AdS_3}\times X$ is 
\begin{equation}
{\rm Sym}^N(\mathbb{R}_{\mathcal{Q}}\times \text{3 fermions}\times\beta\gamma\text{ ghosts}\times X)+\int \sigma_{2,\tilde\alpha,\bar{\tilde{\alpha}}}\,,
\end{equation}
where $\tilde\alpha,\bar{\tilde{\alpha}}$ denote collectively the linear dilaton momentum and other labellings. Effectively, the $\beta\gamma$ ghosts\footnote{These are the usual superconformal ghosts and should not be confused with the $\beta\gamma$ in the Wakimoto representation.} kill two of the fermions and a tentative candidate for a dual CFT is thus
\begin{equation}
{\rm Sym}^N(\mathbb{R}_{\mathcal{Q}}\times \text{1 fermion}\times X)+\int \sigma_{2,\tilde\alpha,\bar{\tilde{\alpha}}}\,.
\end{equation}
Specializing to the case where $X={\rm S^3}\times\tilde X$ with $\tilde X=K3$ or $\mathbb{T}^4$, we see that the dual CFT can be written as
\begin{equation}
{\rm Sym}^N(\mathbb{R}_{\mathcal{Q}}\times \text{4 fermions}\times \mathfrak{su}(2)_{k-2}\times \tilde X)+\int \sigma_{2,\tilde\alpha,\bar{\tilde{\alpha}}}\,,
\end{equation}
where we exmphasize that $k$ in the $\mathfrak{su}(2)$ level is the same as the supersymmetric level of $\mathfrak{sl}(2,\mathbb{R})$ as is required by string criticality. This is the proposal of \cite{Eberhardt:2021vsx}. However, it remains to be seen precisely how the $\beta\gamma$ ghost cancellation works in any generic correlators. This is currently under investigation \cite{Sriprachyakul:2024def}.

\paragraph{\boldmath String theory on $G^{\,\mathbb{C}}/G$ coset spaces:}

The fundamental tool used throughout this paper was the realization of string theory on Euclidean $\text{AdS}_3$ in terms of the Wakimoto free fields $\Phi,\gamma,\bar{\gamma},\beta,\bar{\beta}$. A more general class of string backgrounds admitting analogous free field realizations are the coset spaces $G^{\,\mathbb{C}}/G$, where $G$ is some real Lie group and $G^{\,\mathbb{C}}$ is its complexification \cite{Gawedzki:1991yu}. If $G=\text{SU}(2)$, then $G^{\,\mathbb{C}}/G$ is simply Euclidean $\text{AdS}_3$. It would be interesting to explore whether the techniques developed in this paper would allow one to calculate correlation functions in such coset spaces for more general groups $G$ and, if so, whether there is some `holographic' relation to a QFT on the asymptotic boundary.

\paragraph{Relation to Gromov-Witten theory:}

In the calculation of the worldsheet path integral, the computation of the $\beta\gamma$ system reduced to an integral over the moduli space of holomorphic maps $\gamma:\Sigma\to B$ of degree $N$ with $B$ the boundary of $\text{AdS}_3$. One can take the integration over such holomorphic maps and the integral over the worldsheet moduli into an integral over the Kontsevich moduli space $\mathcal{M}_{g,n}(B,N)$ of holomorphic curves in $B$ of degree $N$ with $n$ marked points. This is a moduli space of (virtual) dimension
\begin{equation}
\begin{split}
\text{dim}(\mathcal{M}_{g,n}(B,N))&=n+2g-2+N\int_{B}c_1(B)\\
&=n+2g-2+N(2-2G)\,,
\end{split}
\end{equation}
where $G$ is the genus of the boundary $B$. From this point of view, we can think of the delta function operators introduced in Section \ref{sec:ads3-strings}, namely
\begin{equation}
\delta^{(2)}_{w_i}(\gamma(z_i)-x_i)
\end{equation}
as distributions on the moduli space $\mathcal{M}_{g,n}(B,N)$ which localize to codimension $w_i$ subvarieties. The computation of spectrally-flowed correlators can thus be rephrased in terms of intersection numbers of subvarieties on the Kontsevich moduli space. Mathematically, the study of these intersection numbers is the realm of Gromov-Witten theory. Given that Gromov-Witten theory with a curve as a target is known to be equivalent to Hurwitz theory \cite{Okounkov:2002cja}, understanding the relation between $\text{AdS}_3$ string theory and Gromov-Witten theory could help elucidate the relationship between $\text{AdS}_3$ string theory and symmetric orbifolds.\footnote{We suspect the that work of \cite{Frenkel:2006fy,Frenkel:2008vz,janda2023gromovwitten,Lerche:2023wkj} will be instrumental to understanding this relationship in more detail.}

A topological version of this idea has recently been made concrete \cite{Lerche:2023wkj}, where the author argued that the correspondence between the quantum cohomology of the Hilbert scheme $\text{Hilb}(\text{K}3)$ and Gromov-Witten theory on $\mathbb{CP}^1\times\text{K}3$ corresponds to a topological subsector of the $\text{AdS}_3/\text{CFT}_2$ correspondence. It would be nice to see how this mathematical result fits into the language of $\text{AdS}_3$ string theory.

\acknowledgments
We thank Bruno Balthazar, Andrea Dei, Lorenz Eberhardt, Matthias Gaberdiel, Yasuaki Hikida, Edward Mazenc, Volker Schomerus, Sean Seet and David Skinner for useful discussions. We especially thank Andrea Dei, Matthias Gaberdiel, and Nicolas Kovensky for helpful comments on an early version of the draft. The work of BK is supported by STFC consolidated grants ST/T000694/1 and ST/X000664/1. The work of VS is supported by the NCCR SwissMAP that is also supported by the Swiss National Science Foundation.

\appendix

\section{\boldmath The \texorpdfstring{$\beta\gamma$}{beta-gamma} path integral}\label{betagammaappen}

In this appendix, we derive the expression \eqref{eq:jacobian-main-text} for the Jacobian of the $\beta\gamma$ path integral under the change of variables \eqref{eq:moduli-space-coordinate-change}, which is necessary in obtaining a matching between the worldsheet and spacetime correlation functions. The computation involves a complicated row reduction on large matrices, which we show to end in a finite number of steps. The end result is a simple epxression in terms of analytic data of the corresponding branched covering maps. We work purely at genus $g=0$ and only focus on the left-moving sector of the theory for simplicity.

\subsection{Deriving the measure}\label{sec:deriving path integral measure}

In this subsection, we give another derivation of the $\gamma$ path integral measure. We start by noticing that the path integral of $\beta,\gamma$ is over the space of functions that depends only on $z$ (and not on $\bar z$) and have pole singularities at isolated points. The most general form of such a function is 
\begin{equation}
\begin{aligned}
\gamma(z)=\sum_{n>0, w\in\mathbb{C}}\frac{c_{n}^w}{(z-w)^n}+\sum_{n\geq 0}b_nz^n\,,
\label{g=0expansion}
\end{aligned}
\end{equation}
where the sum over $w$ runs through all the locations of poles of $\gamma(z)$. Let us simplify our discussion further by assuming that $z=\infty$ is a regular point.\footnote{If $\gamma$ has a pole at $z=\infty$, we can simply perform a M\"obius transformation $\gamma\to (a\gamma+b)/(c\gamma+d)$ so that $\gamma(\infty)$ becomes finite.} This means that we restrict ourselves to the expansion
\begin{equation}
\begin{aligned}
\gamma(z)=\sum_{n>0, w\in\mathbb{C}}\frac{c_{n}^w}{(z-w)^n}+b\,.
\label{}
\end{aligned}
\end{equation}
Hence, instead of taking the path integral measure to be 
\begin{equation}
\begin{aligned}
\mathcal{D}\gamma=\prod_{z\in\mathbb{C}}\mathrm{d}\gamma(z),
\label{}
\end{aligned}
\end{equation}
we should instead change the variable so that the path integral measure is
\begin{equation}
\begin{aligned}
\mathcal{D}'\gamma=\left(\prod_{n>0,w\in\mathbb{C}}\mathrm{d}c^w_n\right)\mathrm{d}b\,.
\label{}
\end{aligned}
\end{equation}
Indeed, the Jacobian of this change of variable is the determinant of the matrix with components\footnote{Here, the rows are labelled by $z$ and the columns are labelled by $w,n$.}
\begin{equation}
\begin{aligned}
\begin{pmatrix}
\displaystyle\frac{\mathrm{d}\gamma(z)}{\mathrm{d}c^w_n} & & \displaystyle\frac{\mathrm{d}\gamma(z)}{\mathrm{d}b}
\end{pmatrix}
=\begin{pmatrix}
\displaystyle\frac{1}{(z-w)^n} & & \displaystyle1
\end{pmatrix}\,.
\label{}
\end{aligned}
\end{equation}
And since this does not depend on $c^w_n$ or $b$, it is just an overall constant which can be absorbed into the normalization of the path integral. Thus the $\beta\gamma$ path integral with the insertions of $V^{w_i}_{m_i,j_i}$ and $D$ reads
\begin{equation}
\begin{aligned}
&\int\mathcal{D}\beta\left(\prod_{n>0,w\in\mathbb{C}}\mathrm{d}c^w_n\right)\mathrm{d}b\,e^{-S[\beta,\gamma]}\prod_{a=1}^{N}\left(\oint_{\lambda_a}\gamma\right)^{-(k-1)}\delta(\beta(\lambda_a))\\
&\hspace{3cm}\times\prod_{i=1}^{n}\left(\frac{\partial^{w_i}(\gamma(z_i)-x_i)}{w_i!}\right)^{-m_i-j_i}\delta_{w_i}(\gamma(z_i)-x_i)\,.
\label{3.6}
\end{aligned}
\end{equation}
By writing $\delta(\beta)$ using the (formal) identity
\begin{equation}
\delta(\beta(\lambda_a))=\int\frac{\mathrm{d}\xi_a}{2\pi}e^{i\beta(\lambda_a)\xi_a}\,
\end{equation}
and performing the integration over $\mathcal{D}\beta(z)$, we obtain
\begin{equation}
\mathcal{D}\beta\exp\left( -
\frac{1}{2\pi}\int_{\Sigma}\beta\left(\overline{\partial}\gamma-\sum_a2\pi i\xi_a\delta^{(2)}(z,\lambda_a)\right) \right)=\delta\left(\bar\partial\gamma-\sum_a2\pi i\xi_a\delta^{(2)}(z,\lambda_a)\right)\,
\label{dfinal}
\end{equation}
where the RHS is a Dirac functional. This Dirac functional restricts $\gamma(z)$ to have $N$ simple poles at $\lambda_a$ with residue $i\xi_a$. Hence, the path integral \eqref{3.6} becomes
\begin{equation}
\begin{aligned}
&\int\left(\prod_{n>0,w\in\mathbb{C}}\mathrm{d}c^w_n\right)\mathrm{d}b\prod_{a=1}^N\mathrm{d}\xi_a\,\prod_{n>2,w\in\mathbb{C}}\delta(c^w_n)\prod_{w\neq\lambda_a}\delta(c^w_1)\prod_{w=\lambda_a}\delta(c^w_1-i\xi_a)\\
&\hspace{2cm}\times\prod_{a=1}^{N}\left(\oint_{\lambda_a}\gamma\right)^{-(k-1)}\prod_{i=1}^{n}\left(\frac{\partial^{w_i}(\gamma(z_i)-x_i)}{w_i!}\right)^{-m_i-j_i}\delta_{w_i}(\gamma(z_i)-x_i)\,.
\label{}
\end{aligned}
\end{equation}
Notice that in the last line, we have have rewritten the Dirac functional \eqref{dfinal} as the product of delta functions imposing the equivalent conditions on the coefficients $c^w_n$. Performing the integrals for $c^w_n$, we obtain
\begin{equation}
\begin{aligned}
&\int\left(\prod_{a=1}^N\mathrm{d}\xi_a\right)\mathrm{d}b\,\prod_{a=1}^{N}\left(\oint_{\lambda_a}\gamma\right)^{-(k-1)}\prod_{i=1}^{n}\left(\frac{\partial^{w_i}(\gamma(z_i)-x_i)}{w_i!}\right)^{-m_i-j_i}\delta_{w_i}(\gamma(z_i)-x_i)\,.
\label{}
\end{aligned}
\end{equation}
Redefining $c^\gamma_a:=\xi_a=\text{Res}_{z=\lambda_a}\gamma(z)$ , we have
\begin{tcolorbox}
\begin{equation}
\begin{aligned}
\int\left(\prod_{a=1}^N\mathrm{d}c^\gamma_a\right)\mathrm{d}b\,\prod_{a=1}^{N}\left(\oint_{\lambda_a}\gamma\right)^{-(k-1)}\prod_{i=1}^{n}\left(\frac{\partial^{w_i}(\gamma(z_i)-x_i)}{w_i!}\right)^{-m_i-j_i}\delta_{w_i}(\gamma(z_i)-x_i)\,.
\label{corrpathint}
\end{aligned}
\end{equation}
\end{tcolorbox}
\noindent We propose that this is the correct measure for the path integral over $\gamma$, once we have restricted the integral over holomorphic functions. Taking into account right-moving degrees of freedom, we obtain the form \eqref{finalgammapathint} of the path integral used in the main text to match worldsheet correlators to $\text{CFT}_2$ ones.

\subsection[Computing the \texorpdfstring{$\beta\gamma$}{beta-gamma} integral Jacobian]{\boldmath Computing the \texorpdfstring{$\beta\gamma$}{beta-gamma} integral Jacobian}
In this section we describe in details how we obtain various Jacobians in the main text. We note here that we will generally ignore any minus signs and phases since we eventually are interested in amplitudes which are modulus squares of the quantity we are computing in this paper. As a warm up, we consider the simplest case and build our way up to the most general case.

\subsubsection*{The simplest case}

To start, let us consider the case of 3-point function whose $\text{SL}(2,\mathbb{R})$ spins are chosen so that the number of extra branch points is $m=0$, see \eqref{jconstraint}. In this case, there are no worldsheet moduli and the $\gamma$ field takes the form
\begin{equation}
\begin{aligned}
\gamma(z)=\sum_{a=1}^N\frac{c^\gamma_a}{z-\lambda_a}+b,
\label{a22}
\end{aligned}
\end{equation}
where we have $2N+1=\sum_iw_i$. The Jacobian we want to compute is
\begin{equation}
\begin{aligned}
\mathcal{J}=&\left| \frac{\partial}{\partial b}\left(\frac{\partial^{l_i}\gamma(z_i)}{l_i!}\right)~~\frac{\partial}{\partial c^\gamma_a}\left(\frac{\partial^{l_i}\gamma(z_i)}{l_i!}\right) ~~\frac{\partial}{\partial \lambda_a}\left(\frac{\partial^{l_i}\gamma(z_i)}{l_i!}\right)\right|\,,
\label{}
\end{aligned}
\end{equation}
where $i$ runs from 1 to 3.
Plugging eq.\eqref{a22} into the Jacobian, we can write $\mathcal{J}$ as the determinant
\begin{equation}
\begin{aligned}
\mathcal{J}
=&\begin{vmatrix}
1 & \frac{1}{z_1-\lambda_1} & ... &\frac{1}{z_1-\lambda_N} & \frac{c^\gamma_1}{(z_1-\lambda_1)^2} & ... & \frac{c^\gamma_N}{(z_1-\lambda_N)^2}\\
0 & \frac{1}{(z_1-\lambda_1)^2} & ... &\frac{1}{(z_1-\lambda_N)^2} & \frac{2c^\gamma_1}{(z_1-\lambda_1)^3} & ... & \frac{2c^\gamma_N}{(z_1-\lambda_N)^3}\\
\vdots & \vdots & ... &\vdots & \vdots & ... & \vdots\\
0 & \frac{1}{(z_1-\lambda_1)^{w_1}} & ... &\frac{1}{(z_1-\lambda_N)^{w_1}} & \frac{w_1c^\gamma_1}{(z_1-\lambda_1)^{w_1+1}} & ... & \frac{w_1c^\gamma_N}{(z_1-\lambda_N)^{w_1+1}}\\
1 & \frac{1}{z_2-\lambda_1} & ... &\frac{1}{z_2-\lambda_N} & \frac{c^\gamma_1}{(z_2-\lambda_1)^2} & ... & \frac{c^\gamma_N}{(z_2-\lambda_N)^2}\\
0 & \frac{1}{(z_2-\lambda_1)^2} & ... &\frac{1}{(z_2-\lambda_N)^2} & \frac{2c^\gamma_1}{(z_2-\lambda_1)^3} & ... & \frac{2c^\gamma_N}{(z_2-\lambda_N)^3}\\
\vdots & \vdots & ... &\vdots & \vdots & ... & \vdots\\
0 & \frac{1}{(z_2-\lambda_1)^{w_2}} & ... &\frac{1}{(z_2-\lambda_N)^{w_2}} & \frac{w_2c^\gamma_1}{(z_2-\lambda_1)^{w_2+1}} & ... & \frac{w_2c^\gamma_N}{(z_2-\lambda_N)^{w_2+1}}\\
1 & \frac{1}{z_3-\lambda_1} & ... &\frac{1}{z_3-\lambda_N} & \frac{c^\gamma_1}{(z_3-\lambda_1)^2} & ... & \frac{c^\gamma_N}{(z_3-\lambda_N)^2}\\
0 & \frac{1}{(z_3-\lambda_1)^2} & ... &\frac{1}{(z_3-\lambda_N)^2} & \frac{2c^\gamma_1}{(z_3-\lambda_1)^3} & ... & \frac{2c^\gamma_N}{(z_3-\lambda_N)^3}\\
\vdots & \vdots & ... &\vdots & \vdots & ... & \vdots\\
0 & \frac{1}{(z_3-\lambda_1)^{w_3}} & ... &\frac{1}{(z_3-\lambda_N)^{w_3}} & \frac{w_3c^\gamma_1}{(z_3-\lambda_1)^{w_3+1}} & ... & \frac{w_3c^\gamma_N}{(z_3-\lambda_N)^{w_3+1}}\\
\end{vmatrix}\,.\\
\label{}
\end{aligned}
\end{equation}
We will now compute this determinant and we divide the calculation into various simple steps.

\paragraph{Step 1:} 
First, we factor out the common factors of $c^\gamma_a$ in the last $N$ columns. Then, we get rid of the 1's in the first column by subtracting the first row from the $(w_1+1)^{th}$ and $(w_1+w_2+1)^{th}$ rows. We obtain
\begin{equation}
z_{12}z_{13}\prod_{a=1}^Nc^\gamma_a\\
\begin{vmatrix}
1 & ... & \frac{1}{z_1-\lambda_a} & ... & \frac{1}{(z_1-\lambda_a)^2} & ... \\
\vdots & \vdots & \vdots &\vdots & \vdots & \vdots\\
0 & ... & \frac{1}{(z_1-\lambda_a)^m} & ... & \frac{m}{(z_1-\lambda_a)^{m+1}} & ...\\
\vdots & \vdots & ... &\vdots & \vdots & \vdots\\
0 & ... & \frac{1}{(z_1-\lambda_a)^{w_1}} & ... & \frac{w_1}{(z_1-\lambda_a)^{w_1+1}} & ... \\
0 & ... & \frac{1}{(z_1-\lambda_a)(z_2-\lambda_a)} & ... & \frac{1}{(z_2-\lambda_a)^2(z_1-\lambda_a)}+\frac{1}{(z_2-\lambda_a)(z_1-\lambda_a)^2} & ... \\
0 & ... & \frac{1}{(z_2-\lambda_a)^2} & ... & \frac{2}{(z_2-\lambda_a)^3} & ... \\
\vdots & \vdots & \vdots &\vdots & \vdots & \vdots\\
0 & ... & \frac{1}{(z_2-\lambda_a)^{w_2}} & ... & \frac{w_2}{(z_2-\lambda_a)^{w_2+1}} & ...\\
0 & ... & \frac{1}{(z_1-\lambda_a)(z_3-\lambda_a)} & ... & \frac{1}{(z_3-\lambda_a)^2(z_1-\lambda_a)}+\frac{1}{(z_3-\lambda_a)(z_1-\lambda_a)^2} & ...\\
0 & ... & \frac{1}{(z_3-\lambda_a)^2} & ... & \frac{2}{(z_3-\lambda_a)^3} & ...\\
\vdots & \vdots & \vdots &\vdots & \vdots & \vdots\\
0 & ... & \frac{1}{(z_3-\lambda_a)^{w_3}} & ... & \frac{w_3}{(z_3-\lambda_a)^{w_3+1}} & ...\\
\end{vmatrix}.
\label{}
\end{equation}
\paragraph{Step 2:}
Next, we subtract the row that contains $(\tfrac{1}{(z_1-\lambda_a)(z_2-\lambda_a)},\tfrac{1}{(z_2-\lambda_a)^2(z_1-\lambda_a)}+\tfrac{1}{(z_2-\lambda_a)(z_1-\lambda_a)^2})$ from the row that contains $(\tfrac{1}{(z_2-\lambda_a)^2},\tfrac{2}{(z_2-\lambda_a)^3})$. We then use the resulting expression to subtract from the row that contains $(\tfrac{1}{(z_2-\lambda_a)^3},\tfrac{3}{(z_2-\lambda_a)^4})$ and we repeat the process until we subtract from the row that contains $\tfrac{1}{(z_2-\lambda_a)^{w_2}}$. We then do the same for the rows that depend on $z_3$. The result is
\begin{equation}
z_{12}^{w_2}z_{13}^{w_3}\prod_{a=1}^Nc^\gamma_a\\
\begin{vmatrix}
1 & ... & \frac{1}{z_1-\lambda_a} & ... & \frac{1}{(z_1-\lambda_a)^2} & ...\\
0 & ... & \frac{1}{(z_1-\lambda_a)^2} & ... & \frac{2}{(z_1-\lambda_a)^3} & ... \\
\vdots & \vdots & \vdots &\vdots & \vdots & \vdots \\
0 & ... & \frac{1}{(z_1-\lambda_a)^{w_1}} & ... & \frac{w_1}{(z_1-\lambda_a)^{w_1+1}} & ... \\
0 & ...& \frac{1}{(z_1-\lambda_a)(z_2-\lambda_a)} & ... & \frac{1}{(z_2-\lambda_a)^2(z_1-\lambda_a)}+\frac{1}{(z_2-\lambda_a)(z_1-\lambda_a)^2} & ...\\
\vdots & \vdots & \vdots &\vdots & \vdots & \vdots\\
0 & ... & \frac{1}{(z_1-\lambda_a)(z_2-\lambda_a)^m} & ... & \frac{m}{(z_1-\lambda_a)(z_2-\lambda_a)^{m+1}}+\frac{1}{(z_1-\lambda_a)^2(z_2-\lambda_a)^m} & ... \\
\vdots & \vdots & \vdots &\vdots & \vdots & \vdots \\
0 & ... & \frac{1}{(z_1-\lambda_a)(z_2-\lambda_a)^{w_2}} & ... & \frac{w_2}{(z_1-\lambda_a)(z_2-\lambda_a)^{w_2+1}}+\frac{1}{(z_1-\lambda_a)^2(z_2-\lambda_a)^{w_2}} & ...\\
0 & ... & \frac{1}{(z_1-\lambda_a)(z_3-\lambda_a)} & ... & \frac{1}{(z_3-\lambda_a)^2(z_1-\lambda_a)}+\frac{1}{(z_3-\lambda_a)(z_1-\lambda_a)^2} & ...\\
\vdots & \vdots & \vdots &\vdots & \vdots & \vdots \\
0 & ... & \frac{1}{(z_1-\lambda_a)(z_3-\lambda_a)^m} & ... & \frac{m}{(z_1-\lambda_a)(z_3-\lambda_a)^{m+1}}+\frac{1}{(z_1-\lambda_a)^2(z_3-\lambda_a)^m} & ...\\
\vdots & \vdots & \vdots &\vdots & \vdots & \vdots \\
0 & ... & \frac{1}{(z_1-\lambda_a)(z_3-\lambda_a)^{w_3}} & ... & \frac{w_3}{(z_1-\lambda_a)(z_3-\lambda_a)^{w_3+1}}+\frac{1}{(z_1-\lambda_a)^2(z_3-\lambda_a)^{w_3}} & ...\\
\end{vmatrix}.
\label{}
\end{equation}
\paragraph{Step 3:}
We then factor out the common factor $\tfrac{1}{(z_1-\lambda_a)}$, obtaining
\begin{equation}
z_{12}^{w_2}z_{13}^{w_3}\prod_{a=1}^N\frac{c^\gamma_a}{z_1-\lambda_a}\\
\begin{vmatrix}
1 & ... & \frac{1}{z_1-\lambda_a} & ... & \frac{1}{(z_1-\lambda_a)} & ... \\
0 & ... & \frac{1}{(z_1-\lambda_a)^2} & ... & \frac{2}{(z_1-\lambda_a)^2} & ...\\
\vdots & \vdots & \vdots &\vdots & \vdots & \vdots\\
0 & ... & \frac{1}{(z_1-\lambda_a)^{w_1}} & ... & \frac{w_1}{(z_1-\lambda_a)^{w_1}} & ...\\
0 & ... & \frac{1}{(z_1-\lambda_a)(z_2-\lambda_a)} & ... & \frac{1}{(z_2-\lambda_a)^2}+\frac{1}{(z_2-\lambda_a)(z_1-\lambda_a)} & ...\\
\vdots & \vdots & \vdots &\vdots & \vdots & \vdots\\
0 & ... & \frac{1}{(z_1-\lambda_a)(z_2-\lambda_a)^m} & ... & \frac{m}{(z_2-\lambda_a)^{m+1}}+\frac{1}{(z_1-\lambda_a)(z_2-\lambda_a)^m} & ...\\
\vdots & \vdots & \vdots &\vdots & \vdots & \vdots\\
0 & ... & \frac{1}{(z_1-\lambda_a)(z_2-\lambda_a)^{w_2}} & ... & \frac{w_2}{(z_2-\lambda_a)^{w_2+1}}+\frac{1}{(z_1-\lambda_a)(z_2-\lambda_a)^{w_2}} & ...\\
0 & ... & \frac{1}{(z_1-\lambda_a)(z_3-\lambda_a)} & ... & \frac{1}{(z_3-\lambda_a)^2}+\frac{1}{(z_3-\lambda_a)(z_1-\lambda_a)} & ...\\
\vdots & \vdots & \vdots &\vdots & \vdots & \vdots\\
0 & ... & \frac{1}{(z_1-\lambda_a)(z_3-\lambda_a)^m} & ... & \frac{m}{(z_3-\lambda_a)^{m+1}}+\frac{1}{(z_1-\lambda_a)(z_3-\lambda_a)^m} & ...\\
\vdots & \vdots & \vdots &\vdots & \vdots & \vdots\\
0 & ... & \frac{1}{(z_1-\lambda_a)(z_3-\lambda_a)^{w_3}} & ... & \frac{w_3}{(z_3-\lambda_a)^{w_3+1}}+\frac{1}{(z_1-\lambda_a)(z_3-\lambda_a)^{w_3}} & ...\\
\end{vmatrix}\,.
\label{a8}
\end{equation}
\paragraph{Step 4:}
We now subtract the columns shown in \eqref{a8} to obtain
\begin{equation}
z_{12}^{w_2}z_{13}^{w_3}\prod_{a=1}^N\frac{c^\gamma_a}{z_1-\lambda_a}
\begin{vmatrix}
1 & \frac{1}{z_1-\lambda_1} & ... &\frac{1}{z_1-\lambda_N} & 0 & ... & 0\\
0 & \frac{1}{(z_1-\lambda_1)^2} & ... &\frac{1}{(z_1-\lambda_N)^2} & \frac{1}{(z_1-\lambda_1)^2} & ... & \frac{1}{(z_1-\lambda_N)^2}\\
\vdots & \vdots & ... &\vdots & \vdots & ... & \vdots\\
0 & \frac{1}{(z_1-\lambda_1)^m} & ... &\frac{1}{(z_1-\lambda_N)^m} & \frac{m-1}{(z_1-\lambda_1)^m} & ... & \frac{m-1}{(z_1-\lambda_N)^m}\\
\vdots & \vdots & ... &\vdots & \vdots & ... & \vdots\\
0 & \frac{1}{(z_1-\lambda_1)^{w_1}} & ... &\frac{1}{(z_1-\lambda_N)^{w_1}} & \frac{w_1-1}{(z_1-\lambda_1)^{w_1}} & ... & \frac{w_1-1}{(z_1-\lambda_N)^{w_1}}\\
0 & \frac{1}{(z_1-\lambda_1)(z_2-\lambda_1)} & ... &\frac{1}{(z_1-\lambda_N)(z_2-\lambda_N)} & \frac{1}{(z_2-\lambda_1)^2} & ... & \frac{1}{(z_2-\lambda_N)^2}\\
\vdots & \vdots & ... &\vdots & \vdots & ... & \vdots\\
0 & \frac{1}{(z_1-\lambda_1)(z_2-\lambda_1)^m} & ... &\frac{1}{(z_1-\lambda_N)(z_2-\lambda_N)^m} & \frac{m}{(z_2-\lambda_1)^{m+1}} & ... & \frac{m}{(z_2-\lambda_N)^{m+1}}\\
\vdots & \vdots & ... &\vdots & \vdots & ... & \vdots\\
0 & \frac{1}{(z_1-\lambda_1)(z_2-\lambda_1)^{w_2}} & ... &\frac{1}{(z_1-\lambda_N)(z_2-\lambda_N)^{w_2}} & \frac{w_2}{(z_2-\lambda_1)^{w_2+1}} & ... & \frac{w_2}{(z_2-\lambda_N)^{w_2+1}}\\
0 & \frac{1}{(z_1-\lambda_1)(z_3-\lambda_1)} & ... &\frac{1}{(z_1-\lambda_N)(z_3-\lambda_N)} & \frac{1}{(z_3-\lambda_1)^2} & ... & \frac{1}{(z_3-\lambda_N)^2}\\
\vdots & \vdots & ... &\vdots & \vdots & ... & \vdots\\
0 & \frac{1}{(z_1-\lambda_1)(z_3-\lambda_1)^m} & ... &\frac{1}{(z_1-\lambda_N)(z_3-\lambda_N)^m} & \frac{m}{(z_3-\lambda_1)^{m+1}} & ... & \frac{m}{(z_3-\lambda_N)^{m+1}}\\
\vdots & \vdots & ... &\vdots & \vdots & ... & \vdots\\
0 & \frac{1}{(z_1-\lambda_1)(z_3-\lambda_1)^{w_3}} & ... &\frac{1}{(z_1-\lambda_N)(z_3-\lambda_N)^{w_3}} & \frac{w_3}{(z_3-\lambda_1)^{w_3+1}} & ... & \frac{w_3}{(z_3-\lambda_N)^{w_3+1}}\\
\end{vmatrix}\,.
\label{}
\end{equation}
Factoring out $\tfrac{1}{z_1-\lambda_a}$ from the various rows, we obtain
\begin{equation}
z_{12}^{w_2}z_{13}^{w_3}\prod_{a=1}^N\frac{c^\gamma_a}{(z_1-\lambda_a)^2}
\begin{vmatrix}
\frac{1}{(z_1-\lambda_1)} & ... &\frac{1}{(z_1-\lambda_N)} & \frac{1}{(z_1-\lambda_1)^2} & ... & \frac{1}{(z_1-\lambda_N)^2}\\
\vdots & ... &\vdots & \vdots & ... & \vdots\\
\frac{1}{(z_1-\lambda_1)^{m-1}} & ... &\frac{1}{(z_1-\lambda_N)^{m-1}} & \frac{m-1}{(z_1-\lambda_1)^m} & ... & \frac{m-1}{(z_1-\lambda_N)^m}\\
\vdots & ... &\vdots & \vdots & ... & \vdots\\
\frac{1}{(z_1-\lambda_1)^{w_1-1}} & ... &\frac{1}{(z_1-\lambda_N)^{w_1-1}} & \frac{w_1-1}{(z_1-\lambda_1)^{w_1}} & ... & \frac{w_1-1}{(z_1-\lambda_N)^{w_1}}\\
\frac{1}{(z_2-\lambda_1)} & ... &\frac{1}{(z_2-\lambda_N)} & \frac{1}{(z_2-\lambda_1)^2} & ... & \frac{1}{(z_2-\lambda_N)^2}\\
\vdots & ... &\vdots & \vdots & ... & \vdots\\
\frac{1}{(z_2-\lambda_1)^m} & ... &\frac{1}{(z_2-\lambda_N)^m} & \frac{m}{(z_2-\lambda_1)^{m+1}} & ... & \frac{m}{(z_2-\lambda_N)^{m+1}}\\
\vdots & ... &\vdots & \vdots & ... & \vdots\\
\frac{1}{(z_2-\lambda_1)^{w_2}} & ... &\frac{1}{(z_2-\lambda_N)^{w_2}} & \frac{w_2}{(z_2-\lambda_1)^{w_2+1}} & ... & \frac{w_2}{(z_2-\lambda_N)^{w_2+1}}\\
\frac{1}{(z_3-\lambda_1)} & ... &\frac{1}{(z_3-\lambda_N)} & \frac{1}{(z_3-\lambda_1)^2} & ... & \frac{1}{(z_3-\lambda_N)^2}\\
\vdots & ... &\vdots & \vdots & ... & \vdots\\
\frac{1}{(z_3-\lambda_1)^m} & ... &\frac{1}{(z_3-\lambda_N)^m} & \frac{m}{(z_3-\lambda_1)^{m+1}} & ... & \frac{m}{(z_3-\lambda_N)^{m+1}}\\
\vdots & ... &\vdots & \vdots & ... & \vdots\\
\frac{1}{(z_3-\lambda_1)^{w_3}} & ... &\frac{1}{(z_3-\lambda_N)^{w_3}} & \frac{w_3}{(z_3-\lambda_1)^{w_3+1}} & ... & \frac{w_3}{(z_3-\lambda_N)^{w_3+1}}\\
\end{vmatrix}\,.
\end{equation}
\paragraph{Step 5:}
We now repeat the steps $1)-4)$ above,
\begin{equation}
\begin{aligned}
\mathcal{J}=&\prod_{a=1}^Nc^\gamma_az_{12}^{w_2}z_{13}^{w_3}\prod_{a=1}^N\frac{1}{(z_1-\lambda_a)^2}z_{12}^{w_2}z_{13}^{w_3}\prod_{a=1}^N\frac{1}{(z_1-\lambda_a)^2}\\
&\begin{vmatrix}
1 & ... & 1 & 0 & ... & 0\\
\frac{1}{(z_1-\lambda_1)} & ... &\frac{1}{(z_1-\lambda_N)} & \frac{1}{(z_1-\lambda_1)^2} & ... & \frac{1}{(z_1-\lambda_N)^2}\\
\vdots & ... &\vdots & \vdots & ... & \vdots\\
\frac{1}{(z_1-\lambda_1)^{m-1}} & ... &\frac{1}{(z_1-\lambda_N)^{m-1}} & \frac{m-1}{(z_1-\lambda_1)^m} & ... & \frac{m-1}{(z_1-\lambda_N)^m}\\
\vdots & ... &\vdots & \vdots & ... & \vdots\\
\frac{1}{(z_1-\lambda_1)^{w_1-2}} & ... &\frac{1}{(z_1-\lambda_N)^{w_1-2}} & \frac{w_1-2}{(z_1-\lambda_1)^{w_1-1}} & ... & \frac{w_1-2}{(z_1-\lambda_N)^{w_1-1}}\\
\frac{1}{(z_2-\lambda_1)} & ... &\frac{1}{(z_2-\lambda_N)} & \frac{1}{(z_2-\lambda_1)^2} & ... & \frac{1}{(z_2-\lambda_N)^2}\\
\frac{1}{(z_2-\lambda_1)^2} & ... &\frac{1}{(z_2-\lambda_N)^2} & \frac{2}{(z_2-\lambda_1)^{3}} & ... & \frac{2}{(z_2-\lambda_N)^{3}}\\
\vdots & ... &\vdots & \vdots & ... & \vdots\\
\frac{1}{(z_2-\lambda_1)^m} & ... &\frac{1}{(z_2-\lambda_N)^m} & \frac{m}{(z_2-\lambda_1)^{m+1}} & ... & \frac{m}{(z_2-\lambda_N)^{m+1}}\\
\vdots & ... &\vdots & \vdots & ... & \vdots\\
\frac{1}{(z_2-\lambda_1)^{w_2}} & ... &\frac{1}{(z_2-\lambda_N)^{w_2}} & \frac{w_2}{(z_2-\lambda_1)^{w_2+1}} & ... & \frac{w_2}{(z_2-\lambda_N)^{w_2+1}}\\
\frac{1}{(z_3-\lambda_1)} & ... &\frac{1}{(z_3-\lambda_N)} & \frac{1}{(z_3-\lambda_1)^2} & ... & \frac{1}{(z_3-\lambda_N)^2}\\
\frac{1}{(z_3-\lambda_1)^2} & ... &\frac{1}{(z_3-\lambda_N)^2} & \frac{2}{(z_3-\lambda_1)^{3}} & ... & \frac{2}{(z_3-\lambda_N)^{3}}\\
\vdots & ... &\vdots & \vdots & ... & \vdots\\
\frac{1}{(z_3-\lambda_1)^m} & ... &\frac{1}{(z_3-\lambda_N)^m} & \frac{m}{(z_3-\lambda_1)^{m+1}} & ... & \frac{m}{(z_3-\lambda_N)^{m+1}}\\
\vdots & ... &\vdots & \vdots & ... & \vdots\\
\frac{1}{(z_3-\lambda_1)^{w_3}} & ... &\frac{1}{(z_3-\lambda_N)^{w_3}} & \frac{w_3}{(z_3-\lambda_1)^{w_3+1}} & ... & \frac{w_3}{(z_3-\lambda_N)^{w_3+1}}\\
\end{vmatrix}\,.
\label{}
\end{aligned}
\end{equation}
\paragraph{Step 6:}
The idea now is to subtract so that only the first column has 1 in the first row and other columns have 0 in the first row. We do so by subtracting the first column from the other columns. A generic form of column $a$ ($a\neq1$), after subtractions, is
\begin{equation}
\begin{aligned}
\frac{1}{(z_i-\lambda_a)^m}-\frac{1}{(z_i-\lambda_1)^m}=&-\lambda_{1a}\left( \sum_{l=0}^{m-1}\frac{1}{(z_i-\lambda_1)^{l+1}(z_i-\lambda_a)^{m-l}} \right).
\end{aligned}
\end{equation}
And the determinant becomes
\begin{equation}
\begin{aligned}
&\prod_{a=1}^Nc^\gamma_az_{12}^{w_2}z_{13}^{w_3}\prod_{a=1}^N\frac{1}{(z_1-\lambda_a)^2}z_{12}^{w_2}z_{13}^{w_3}\prod_{a=1}^N\frac{1}{(z_1-\lambda_a)^2}\prod_{a=2}^N(-\lambda_{1a})\\
&\begin{vmatrix}
\vdots & ... &\vdots & \vdots & ... & \vdots\\
\sum_{l=0}^{m_i-2}\frac{1}{(z_i-\lambda_1)^{l+1}(z_i-\lambda_2)^{m_i-l-1}} & ... &\sum_{l=0}^{m_i-2}\frac{1}{(z_i-\lambda_1)^{l+1}(z_i-\lambda_N)^{m_i-l-1}} & \frac{m_i-1}{(z_i-\lambda_1)^{m_i}} & ... & \frac{m_i-1}{(z_i-\lambda_N)^{m_i}}\\
\vdots & ... &\vdots & \vdots & ... & \vdots\\
\end{vmatrix}.
\label{}
\end{aligned}
\end{equation}
Here, $m_1=2,\ldots,w_1-1$ and $m_{2,3}=2,\ldots,w_{2,3}+1$.
We then subtract the combination $\sum_{l=0}^{m_i-2}\frac{1}{(z_i-\lambda_1)^{l+1}(z_i-\lambda_a)^{m_i-l-1}}$ from $\tfrac{m_i-1}{(z_i-\lambda_a)^{m_i}}$ to obtain
\begin{equation}
\begin{aligned}
&\frac{m_i-1}{(z_i-\lambda_a)^{m_i}}-\sum_{l=0}^{m_i-2}\frac{1}{(z_i-\lambda_1)^{l+1}(z_i-\lambda_a)^{m_i-l-1}}\\
&=-\lambda_{1a}\sum_{l=0}^{m_i-2}\sum_{k=0}^l\frac{1}{(z_i-\lambda_1)^{k+1}(z_i-\lambda_a)^{m_i-k}}=-\lambda_{1a}\sum_{k=0}^{m_i-2}\frac{m_i-1-k}{(z_i-\lambda_1)^{k+1}(z_i-\lambda_a)^{m_i-k}}.
\label{a35}
\end{aligned}
\end{equation}
We then notice that the difference between the $m_i^{th}\times(z_i-\lambda_1)^{-1}$ and $(m_i+1)^{th}$ rows can be written as
\begin{equation}
\begin{aligned}
&\left( \sum_{l=0}^{m_i-2}\frac{1}{(z_i-\lambda_1)^{l+1}(z_i-\lambda_a)^{m_i-l-1}} \right)-\frac{1}{(z_i-\lambda_1)}\left( \sum_{l=0}^{m_i-3}\frac{1}{(z_i-\lambda_1)^{l+1}(z_i-\lambda_a)^{m_i-2-1}} \right)\\
&=\frac{1}{(z_i-\lambda_1)(z_i-\lambda_a)^{m_i-1}}\ ,\\
&\sum_{k=0}^{m_i-2}\frac{m_i-1-k}{(z_i-\lambda_1)^{k+1}(z_i-\lambda_a)^{m_i-k}}-\frac{1}{(z_i-\lambda_1)}\left( \sum_{k=0}^{m_i-3}\frac{m_i-2-k}{(z_i-\lambda_1)^{k+1}(z_i-\lambda_a)^{m_i-k-1}} \right)\\
&=\frac{m_i-1}{(z_i-\lambda_1)(z_i-\lambda_a)^{m_i}},
\label{a36}
\end{aligned}
\end{equation}
and
\begin{equation}
\begin{aligned}
\frac{m_i-1}{(z_i-\lambda_1)^{m_i}}-\frac{m_i-2}{(z_i-\lambda_1)^{m_i}}=\frac{1}{(z_i-\lambda_1)^{m_i}}.
\end{aligned}
\end{equation}
Factoring out the $\tfrac{1}{z_i-\lambda_1}$ then gives
\begin{equation}
\begin{aligned}
&\prod_{a=1}^Nc^\gamma_az_{12}^{w_2}z_{13}^{w_3}\prod_{a=1}^N\frac{1}{(z_1-\lambda_a)^2}z_{12}^{w_2}z_{13}^{w_3}\prod_{a=1}^N\frac{1}{(z_1-\lambda_a)^2}\prod_{a=2}^N(\lambda_{1a})^2\\
&\times\frac{1}{(z_1-\lambda_1)^{w_1-2}}\prod_{i=2}^3\frac{1}{(z_i-\lambda_1)^{w_i}}\,\begin{vmatrix}
\vdots & ... &\vdots & \vdots & ... & \vdots\\
\frac{1}{(z_i-\lambda_1)^{m-1}} & ... &\frac{1}{(z_i-\lambda_N)^{m-1}} & \frac{m-1}{(z_i-\lambda_2)^m} & ... & \frac{m-1}{(z_i-\lambda_N)^m}\\
\vdots & ... &\vdots & \vdots & ... & \vdots\\
\end{vmatrix}\,.
\label{}
\end{aligned}
\end{equation}
Note that maximum value of $m_1$ is $w_1-1$ and of $m_{2,3}$ is $w_{2,3}+1$ while the minimum value of $m_i$ is 2 for all $i$. Note also the fact that there is only one column depending on $\lambda_1$ (which is the first column) and two each which depend on $\lambda_i, i>1$. This agrees with the counting since $w_1-2+w_2+w_3=2N+1-2=2N-1=N+(N-1)$. 
\paragraph{Step 7:}
We now perform steps $1)-6)$. 
The determinant is simplified to
\begin{equation}
\begin{aligned}
&\prod_{a=1}^Nc^\gamma_az_{12}^{3w_2}z_{13}^{3w_3}\prod_{i=1}^3\frac{1}{(z_i-\lambda_1)^{2w_i}}\prod_{a=2}^N\frac{1}{(z_1-\lambda_a)^{6}}\prod_{a=2}^N\lambda_{1a}^4\\
&\hspace{1cm}\times\begin{vmatrix}
\vdots & ... &\vdots & \vdots & ... & \vdots\\
\frac{1}{(z_i-\lambda_2)^{m-1}} & ... &\frac{1}{(z_i-\lambda_N)^{m-1}} & \frac{m-1}{(z_i-\lambda_2)^m} & ... & \frac{m-1}{(z_i-\lambda_N)^m}\\
\vdots & ... &\vdots & \vdots & ... & \vdots\\
\end{vmatrix}\,.
\label{}
\end{aligned}
\end{equation}
Here $m_1$ runs from 
$2$ to $w_1-2$ and $m_{2,3}$ from $2$ to $w_{2,3}+1$. It is then not hard to do the computation inductively from this point on.
The result is
\begin{equation}
\begin{aligned}
\mathcal{J}=&\left(\prod_{a=1}^Nc^\gamma_a\right)\left( 
\prod_{a=1}^N\frac{1}{(z_1-\lambda_a)^{2w_1}(z_2-\lambda_a)^{2w_2}(z_3-\lambda_a)^{2w_3}} \right)\\
&\times\left( \prod_{a<b}(\lambda_a-\lambda_b)^4 \right)z_{12}^{w_2w_1}z_{13}^{w_3w_1}z_{23}^{w_2w_3}.
\label{n=3ans}
\end{aligned}
\end{equation}
Written in terms of covering map data $(C^\gamma,A^\gamma_i)$, we obtain
\begin{equation}
\begin{aligned}
\mathcal{J}=&\prod_{i=1}^3(A^\gamma_i)^{\frac{w_i+1}{2}}(C^\gamma)^{-2}z_{12}z_{13}z_{23}\,.\\
\label{finalans}
\end{aligned}
\end{equation}
This recovers precisely the proposal we had in the main text for this special case. We note one important observation that the steps taken in deriving eq.\eqref{finalans} need not assume that $i$ takes 3 values. As long as the matrix is square, that is $\sum_{i=1}^nw_i=2N+1$, we can always apply the method we described inductively. The corresponding modification is then to allow $i$ to vary from $1,\ldots,n$. This observation will be important below.

\subsubsection*{The more general case}

Let us now drop the assumption about the number of insertions and only assume that the covering maps have no extra branch points. This requires only a minor modification to the $n=3$ case. We have
\begin{equation}
\begin{aligned}
\mathcal{J}=&\left(\prod_{a=1}^Nc^\gamma_a\right)\left( 
\prod_{a,i}\frac{1}{(z_i-\lambda_a)^{2w_i}} \right)\left( \prod_{a<b}(\lambda_a-\lambda_b)^4 \right)\prod_{i<j}z_{ij}^{w_iw_j-1}z_{12}z_{13}z_{23}(C^\gamma)^{n-3}\,.\\
\label{genern}
\end{aligned}
\end{equation}
We start by writing the Jacobian is in a block lower triangular form where the upper left $(2N+1)\times(2N+1)$ part looks like the one in the previous section but with effective branching $w'_i=w_i$ for $i=1,2,3$ and $w'_i=w_i-1$ for $i=4,\ldots,n$. The lower right $(n-3)\times(n-3)$ is then just a diagonal matrix with elements $\partial^{w_i}\gamma(z_i)/(w_i-1)!$ for $i=4,\ldots,n$. Explicitly, we have
\begin{equation}
\mathcal{J}=\begin{vmatrix}
1 & \frac{1}{z_1-\lambda_1} & ... &\frac{1}{z_1-\lambda_N} & \frac{c^\gamma_1}{(z_1-\lambda_1)^2} & ... & \frac{c^\gamma_N}{(z_1-\lambda_N)^2} & 0 & ... & 0\\
0 & \frac{1}{(z_1-\lambda_1)^2} & ... &\frac{1}{(z_1-\lambda_N)^2} & \frac{2c^\gamma_1}{(z_1-\lambda_1)^3} & ... & \frac{2c^\gamma_N}{(z_1-\lambda_N)^3}\\
\vdots & \vdots & ... &\vdots & \vdots & ... & \vdots\\
0 & \frac{1}{(z_1-\lambda_1)^{w_1}} & ... &\frac{1}{(z_1-\lambda_N)^{w_1}} & \frac{w_1c^\gamma_1}{(z_1-\lambda_1)^{w_1+1}} & ... & \frac{w_1c^\gamma_N}{(z_1-\lambda_N)^{w_1+1}}\\
1 & \frac{1}{z_2-\lambda_1} & ... &\frac{1}{z_2-\lambda_N} & \frac{c^\gamma_1}{(z_2-\lambda_1)^2} & ... & \frac{c^\gamma_N}{(z_2-\lambda_N)^2}\\
0 & \frac{1}{(z_2-\lambda_1)^2} & ... &\frac{1}{(z_2-\lambda_N)^2} & \frac{2c^\gamma_1}{(z_2-\lambda_1)^3} & ... & \frac{2c^\gamma_N}{(z_2-\lambda_N)^3}\\
\vdots & \vdots & ... &\vdots & \vdots & ... & \vdots\\
0 & \frac{1}{(z_2-\lambda_1)^{w_2}} & ... &\frac{1}{(z_2-\lambda_N)^{w_2}} & \frac{w_2c^\gamma_1}{(z_2-\lambda_1)^{w_2+1}} & ... & \frac{w_2c^\gamma_N}{(z_2-\lambda_N)^{w_2+1}}\\
1 & \frac{1}{z_3-\lambda_1} & ... &\frac{1}{z_3-\lambda_N} & \frac{c^\gamma_1}{(z_3-\lambda_1)^2} & ... & \frac{c^\gamma_N}{(z_3-\lambda_N)^2}\\
0 & \frac{1}{(z_3-\lambda_1)^2} & ... &\frac{1}{(z_3-\lambda_N)^2} & \frac{2c^\gamma_1}{(z_3-\lambda_1)^3} & ... & \frac{2c^\gamma_N}{(z_3-\lambda_N)^3}\\
\vdots & \vdots & ... &\vdots & \vdots & ... & \vdots & \vdots & \vdots & \vdots\\
0 & \frac{1}{(z_3-\lambda_1)^{w_3}} & ... &\frac{1}{(z_3-\lambda_N)^{w_3}} & \frac{w_3c^\gamma_1}{(z_3-\lambda_1)^{w_3+1}} & ... & \frac{w_3c^\gamma_N}{(z_3-\lambda_N)^{w_3+1}}\\
1 & \frac{1}{z_4-\lambda_1} & ... &\frac{1}{z_4-\lambda_N} & \frac{c^\gamma_1}{(z_4-\lambda_1)^2} & ... & \frac{c^\gamma_N}{(z_4-\lambda_N)^2}\\
0 & \frac{1}{(z_4-\lambda_1)^2} & ... &\frac{1}{(z_4-\lambda_N)^2} & \frac{2c^\gamma_1}{(z_4-\lambda_1)^3} & ... & \frac{2c^\gamma_N}{(z_4-\lambda_N)^3}\\
\vdots & \vdots & ... &\vdots & \vdots & ... & \vdots\\
0 & \frac{1}{(z_4-\lambda_1)^{w_4-1}} & ... &\frac{1}{(z_4-\lambda_N)^{w_4-1}} & \frac{(w_4-1)c^\gamma_1}{(z_4-\lambda_1)^{w_4}} & ... & \frac{(w_4-1)c^\gamma_N}{(z_4-\lambda_N)^{w_4}}\\
\vdots & \vdots & \vdots & \vdots & \vdots & \vdots \\
1 & \frac{1}{z_n-\lambda_1} & ... &\frac{1}{z_n-\lambda_N} & \frac{c^\gamma_1}{(z_n-\lambda_1)^2} & ... & \frac{c^\gamma_N}{(z_n-\lambda_N)^2}\\
0 & \frac{1}{(z_n-\lambda_1)^2} & ... &\frac{1}{(z_n-\lambda_N)^2} & \frac{2c^\gamma_1}{(z_n-\lambda_1)^3} & ... & \frac{2c^\gamma_N}{(z_n-\lambda_N)^3}\\
\vdots & \vdots & ... &\vdots & \vdots & ... & \vdots\\
0 & \frac{1}{(z_n-\lambda_1)^{w_n-1}} & ... &\frac{1}{(z_n-\lambda_N)^{w_n-1}} & \frac{(w_n-1)c^\gamma_1}{(z_n-\lambda_1)^{w_n}} & ... & \frac{(w_n-1)c^\gamma_N}{(z_n-\lambda_N)^{w_n}} & 0 & ... & 0\\
\vdots & \vdots & \vdots & \vdots & \vdots & \vdots & \vdots & \frac{\partial^{w_4}\gamma(z_4)}{(w_4-1)!} & 0 & 0\\
\vdots & \vdots & \vdots & \vdots & \vdots & \vdots & \vdots & ... & \ddots & 0 \\
\vdots & \vdots & \vdots & \vdots & \vdots & \vdots & \vdots & ... & ... & \frac{\partial^{w_n}\gamma(z_n)}{(w_n-1)!}\\
\end{vmatrix}\,.
\end{equation}
Hence, the determinant is the product of the generalization of \eqref{n=3ans} and $\prod_{i=4}^n(w_ia^\gamma_i)$. Indeed, using eq.\eqref{n=3ans} with the effective branchings gives
\begin{equation}
\begin{aligned}
\prod_{a=1}^N(c^\gamma_a)\prod_{a<b}(\lambda_a-\lambda_b)^4\prod_{a,i}\left( \frac{1}{(z_i-\lambda_a)^{2w'_i}} \right)\prod_{i<j}z_{ij}^{w'_iw'_j}\,.
\label{eq:A.33}
\end{aligned}
\end{equation}
And the $\prod_{i=4}^n(w_ia^\gamma_i)$ gives
\begin{equation}
\begin{aligned}
(C^\gamma)^{n-3}\prod_{a,i=4}^n\left( \frac{1}{(z_i-\lambda_a)^{2}} \right)\prod_{i=4}^n\prod_{j\neq i}z_{ij}^{w_j-1}\,.
\label{eq:A.34}
\end{aligned}
\end{equation}
By combining \eqref{eq:A.33} and \eqref{eq:A.34}, this can be shown to reduce to \eqref{genern}.
Writing \eqref{genern} in terms of covering map data therefore gives
\begin{equation}
\begin{aligned}
\mathcal{J}=&\prod_{i=1}^n(A^\gamma_i)^{\frac{w_i+1}{2}}(C^\gamma)^{-2}z_{12}z_{13}z_{23}.
\end{aligned}
\end{equation}

\subsection[The most general case]{\boldmath The most general case}

We can now compute the Jacobian in the most general case of arbitrary $n$-point functions with arbitrary spin, assuming that $m$ defined in equation \eqref{jconstraint} is a non-negative integer. The Jacobian can be computed in essentially the same way as the previous case. Again, the Jacobian is simpler to work out for a 3-point function, this is given by
\begin{equation}
\begin{aligned}
\mathcal{J}=&
\begin{vmatrix}
\frac{\partial}{\partial b}\left(\frac{\partial^{l_i}\gamma(z_i)}{l_i!}\right) & \frac{\partial}{\partial c^\gamma_a}\left(\frac{\partial^{l_i}\gamma(z_i)}{l_i!}\right)  & \frac{\partial}{\partial \lambda_a}\left(\frac{\partial^{l_i}\gamma(z_i)}{l_i!}\right)\\
\frac{\partial}{\partial b}\gamma(\zeta_l) & \frac{\partial}{\partial c^\gamma_a}\gamma(\zeta_l) & \frac{\partial}{\partial \lambda_a}\gamma(\zeta_l)
\end{vmatrix}\\
=&z_{12}^{w_1w_2}z_{13}^{w_1w_3}z_{23}^{w_2w_3}\prod_{a=1}^Nc^\gamma_a 
\prod_{a,i}\frac{1}{(z_i-\lambda_a)^{2w_i}}\prod_{a,l}\frac{1}{(\zeta_l-\lambda_a)^{2}}\\
&\hspace{3cm}\times\prod_{a<b}(\lambda_a-\lambda_b)^4\prod_{i,l}(z_{i}-\zeta_l)^{w_i}\prod_{l<k}\zeta_{lk}\,\,.
\label{}
\end{aligned}
\end{equation}
Indeed, the expression above can be obtained from \eqref{eq:A.33} with the effective branching $w_l=1$ for $l=1,\ldots,m$.
Writing in terms of covering map data, this gives
\begin{equation}
\mathcal{J}=z_{12}z_{13}z_{23}(C^\gamma)^{-2}\prod_{i=1}^3(A^\gamma_i)^{\frac{w_i+1}{2}}\prod_{l=1}^m(A^\gamma_l)^{\frac{1}{2}}\,.
\label{}
\end{equation}
The generalization to $n>3$ is then straightforward by using effective branching explained in the previous section and one can show that the Jacobian takes the form
\begin{tcolorbox}
\begin{equation}
\mathcal{J}=z_{12}z_{13}z_{23}(C^\gamma)^{-2}\prod_{i=1}^3(A^\gamma_i)^{\frac{w_i+1}{2}}\prod_{l=1}^m(A^\gamma_l)^{\frac{1}{2}}\,.
\label{mostgenJ}
\end{equation}
\end{tcolorbox}

\newpage

\bibliography{bibliography}
\bibliographystyle{utphys.bst}

\end{document}